\documentclass{svmult}

\RequirePackage{graphicx} 
\RequirePackage{multicol}
\RequirePackage{makeidx}
\RequirePackage{cite}
\RequirePackage[bottom]{footmisc}

\makeindex

\begin{document} 

\title*{In a book "Tsunami and Nonlinear Waves": Numerical Verification of the Hasselmann equation.} 
\titlerunning{Numerical Verification of the Hasselmann equation.}
\author{Alexander~O.~Korotkevich\inst{1}\and
Andrei~N.~Pushkarev\inst{2,3}\and
Don~Resio\inst{4}\and
Vladimir~E.~Zakharov\inst{5,2,1,3}}
\authorrunning{A.O. Korotkevich, A.N. Pushkarev, D. Resio and V.E. Zakharov}
\institute{Landau Institute for Theoretical Physics RAS, 2, Kosygin Str., Moscow, 119334, Russian Federation
\texttt{kao@landau.ac.ru}
\and Lebedev Physical Institute RAS,53, Leninsky Prosp., GSP-1 Moscow, 119991, Russian Federation
\and Waves and Solitons LLC, 918 W. Windsong Dr., Phoenix, AZ 85045, USA
\texttt{andrei@cox.net}
\and Coastal and Hydraulics Laboratory, U.S. Army Engineer Research and Development Center, Halls Ferry Rd., Vicksburg, MS 39180, USA
\and Department of Mathematics, University of Arizona, Tucson, AZ 85721, USA
\texttt{zakharov@math.arizona.edu}}
\maketitle 
\begin{abstract} 
The purpose of this article is numerical verification of the thory of weak turbulence. 
We performed numerical simulation of an ensemble of nonlinearly interacting free 
gravity waves (swell) by two different methods: solution of primordial dynamical equations
describing potential flow of the ideal fluid with a free surface and, solution of the kinetic
Hasselmann equation, describing the wave ensemble in the framework of the theory of weak turbulence.
Comparison of the results demonstrates pretty good applicability of the weak turbulent 
approach. In both cases we observed effects predicted by this theory: frequency downshift, angular 
spreading as well as formation of Zakharov-Filonenko spectrum $I_{\omega} \sim \omega^{-4}$. To achieve 
quantitative coincidence of the results obtained by different methods one has to 
accomplish the Hasselmann kinetic equation by an empirical dissipation term $S_{diss}$ modeling 
the coherent effects of white-capping. Adding of the standard dissipation terms used in the industrial 
wave predicting model ($WAM$) leads to significant improvement but not resolve the discrepancy 
completely, leaving the question about optimal choice of $S_{diss}$ open. 
 
Numerical modeling of swell evolution in the framework of the dynamical equations is affected by 
the side effect of resonances sparsity taking place due to finite size of the modeling domain. We 
mostly overcame this effect using fine integration grid of $512 \times 4096$ modes. The initial
spectrum peak was located at the wave number $k=300$. Similar conditions can be hardly realized in 
the laboratory wave tanks. One of the results of our article consists in the fact that physical 
processes in finite size laboratory wave tanks and in the ocean are quite different, and the 
results of such laboratory experiments can be applied to modeling of the ocean phenomena with 
extra care. We also present the estimate on the minimum size of the laboratory installation, 
allowing to model open ocean surface wave dynamics. 
\end{abstract} 
 
\section{Introduction.} The theory of weak turbulence is designed for statistical description of 
weakly-nonlinear wave ensembles in dispersive media. The main tool of weak turbulence is 
kinetic equation for squared wave amplitudes, or a system of such equations. Since the 
discovery of the kinetic equation for bosons by Nordheim \cite{Nordheim1928} (see also paper 
by Peierls \cite{Peierls1929}) in the context of solid state physics, this quantum-mechanical 
tool was applied to wide variety of classical problems, including wave turbulence in hydrodynamics, 
plasmas, liquid helium, nonlinear optics, etc. (see monograph by Zakharov, Falkovich and L'vov 
\cite{ZFL1992}). Such kinetic equations have rich families of exact solutions describing 
weak-turbulent Kolmogorov spectra. Also, kinetic equations for waves have self-similar solutions 
describing temporal or spatial evolution of weak -- turbulent spectra. 
 
However, the most remarkable example of weak turbulence is wind-driven sea. The 
kinetic 
equation describing statistically the gravity waves on the surface of ideal 
liquid was 
derived by Hasselmann \cite{Hasselmann1962}. Since this time the Hasselmann 
equation is widely used in physical 
oceanography as foundation for development of wave-prediction models: {\it WAM, 
SWAN} and 
{\it WAVEWATCH} -- it is quite special case between other applications of the theory of 
weak 
turbulence due to the strength of industrial impact. 
 
In spite of tremendous popularity of the Hasselmann equation, its validity and 
applicability 
for description of real wind-driven sea has never been completely proven. It was 
criticized 
by many respected authors, not only in the context of oceanography. There are at 
least two 
reasons why the weak--turbulent theory could fail, or at least be incomplete. 
 
The first reason is presence of the coherent structures. The weak-turbulent 
theory describes 
only weakly-nonlinear resonant processes. Such processes are spatially extended, 
they provide 
weak phase and amplitude correlation on the distances significantly exceeding 
the wave length. 
However, nonlinearity also causes another phenomena, much stronger localized in 
space. Such 
phenomena -- solitons, quasi-solitons and wave collapses are strongly nonlinear 
and cannot be 
described by the kinetic equations. Meanwhile, they could compete with 
weakly-nonlinear resonant 
processes and make comparable or even dominating contribution in the energy, 
momentum and 
wave-action balance. For gravity waves on the fluid surface the most important 
coherent structures 
are white-cappings (or wave-breakings), responsible for essential 
dissipation of wave energy. 
 
The second reason limiting the applicability of the weak-turbulent theory is 
finite size of any 
real physical system. The kinetic equations are derived only for infinite media, 
where the wave 
vector runs continuous {\it D}-dimensional Fourier space. Situation is different 
for the wave 
systems with boundaries, e.g. boxes with periodical or reflective boundary 
conditions. The Fourier 
space of such systems is infinite lattice of discrete eigen-modes. If the 
spacing of the lattice is 
not small enough, or the level of Fourier modes is not big enough, the whole 
physics of nonlinear 
interaction becomes completely different from the continuous case. 
 
For these two reasons verification of the weak turbulent theory is an urgent 
problem, important for 
the whole physics of nonlinear waves. The verification can be done by direct 
numerical simulation of 
the primitive dynamical equations describing wave turbulence in nonlinear 
medium. 
 
So far,the numerical experimentalists tried to check some predictions of the 
weak-turbulent theory, such as 
weak-turbulent Kolmogorov spectra. For the gravity wave turbulence the most 
important is 
Zakharov-Filonenko spectrum $F_\omega\sim\omega^{-4}$ \cite{Zakharov1966}. At 
the moment, this spectrum 
was observed in numerous numerical experiments 
\cite{Pushkarev1996}-\cite{Annenkov2006}. 
 
The attempts of verification of weak turbulent theory through numerical 
simulation of primordial 
dynamical equations has been started with numerical simulation of 2D optical 
turbulence \cite{DNPZ1992}, 
which demonstrated, in particular, co--existence of weak -- turbulent and 
coherent events. 
 
Numerical simulation of 2D turbulence of capillary waves was done in 
\cite{Pushkarev1996}, \cite{Pushkarev1999}, and \cite{Pushkarev2000}. The main 
results of the simulation consisted in observation of classical regime of weak 
turbulence with spectrum $F_\omega\sim\omega^{-19/4}$, and discovery of non-classical 
regime of ``frozen turbulence'', characterized by absence of energy transfer from 
low to high wave-numbers. The classical regime of turbulence was observed 
on the grid of $256 \times 256$ points at relatively high levels of excitation, 
while the ``frozen'' regime was realized at lower levels of excitation, or more coarse 
grids. The effect of ``frozen'' turbulence is explained by sparsity of 3-wave resonance, 
both exact and approximate. The classical regime of turbulence becomes possible due to 
nonlinear shift of the linear frequencies caused by enhanced level of excitation. 
Conclusion has been made that in the reality the turbulence of waves in limited 
systems is practically always the mixture of classical and ``frozen'' regimes.  
 
In fact, the ``frozen'' turbulence is close to $KAM$ regime, when the dynamics 
of turbulence is close to the behavior of integrable system \cite{Pushkarev2000}. 
 
The first attempt to perform modeling of the system of nonlinear waves (swell on 
the surface of deep ocean), solving simultaneously kinetic equation and primordial 
dynamic equations, has been done in the article \cite{Mesoturb2005}. The results of 
this simulation again confirmed ubiquity of the weak-turbulent Zakharov-Filonenko 
asymptotic $\omega^{-4}$ and shown existence of the inverse cascade, but presented 
essentially different scenario of the spectral peak evolution. Detailed analysis 
shown, that even on the grids as fine as $256\times2048$ modes, the energy transport is 
realized mostly by the network of few selected modes -- ``oligarchs'' -- posed in the optimal resonant 
condition. This regime, transitional between weak turbulence and "frozen" turbulence, 
should be typical for wave turbulence in the systems of medium size. It was called 
"mesoscopic turbulence". Similar type of turbulence was observed in \cite{Nazarenko2005}, 
\cite{Nazarenko2006}. 
 
In this article we present the results of new numerical experiments on modeling of swell 
propagation in the framework of both dynamical and kinetic equations, using fine grid 
containing, corresponding to $512 \times 4096$ Fourier modes. We think that our results 
can be considered as first in the world literature direct verification of wave kinetic equation. 
 
One important point should be mentioned. In our experiments we observed not only 
weak turbulence, but also additional nonlinear dissipation of the wave energy, 
which could be identified as the dissipation due to white-capping. To reach 
agreement with dynamic experiments, we had to accomplish the kinetic equation
by a phenomenological dissipation term $S_{diss}$. In this article we examined
dissipation terms used in the industrial wave-prediction models {\it WAM 
Cycle 3} and {WAM cycle 4}. Dissipation term used in {\it WAM Cycle 3} works 
fairly, while $S_{diss}$ used in {WAM Cycle 4} certainly overestimate nonlinear 
dissipation. This fact means that for getting better agreement between dynamic and kinetic 
computations, we need to take into consideration more sophisticated dissipation 
term. 
 
\section{Deterministic and statistic models.} 
 
In the "dynamical" part of our experiment the fluid was described by two 
functions of horizontal variables 
$x,y$ and time $t$: surface elevation $\eta(x,y,t)$ and velocity potential on 
the surface $\psi(x,y,t)$. They 
satisfy the canonical equations \cite{Zakharov1968} 
\begin{equation} 
\label{Hamiltonian_equations} 
\frac{\partial \eta}{\partial t} = \frac{\delta H}{\delta \psi}, \;\;\;\; 
\frac{\partial \psi}{\partial t} = - \frac{\delta H}{\delta \eta}, 
\end{equation} 
Hamiltonian $H$ is presented by the first three terms in expansion on powers of 
nonlinearity $\nabla \eta$ 
\begin{equation} 
\label{Hamiltonian} 
\begin{array}{l} 
\displaystyle 
H = H_0 + H_1 + H_2 + ...,\\ 
\displaystyle 
H_0 = \frac{1}{2}\int\left( g \eta^2 + \psi \hat k  \psi \right) d x d y,\\ 
\displaystyle 
H_1 =  \frac{1}{2}\int\eta\left[ |\nabla \psi|^2 - (\hat k \psi)^2 \right] d x d 
y,\\ 
\displaystyle 
H_2 = \frac{1}{2}\int\eta (\hat k \psi) \left[ \hat k (\eta (\hat k \psi)) + 
\eta\nabla^2\psi \right] d x d y. 
\end{array} 
\end{equation} 
Here $\hat k$ is the linear integral operator 
$\hat k =\sqrt{-\nabla^2}$, defined in Fourier space as 
\begin{equation} 
\hat k \psi_{\vec r} = \frac{1}{2\pi} \int |k| \psi_{\vec k} e^{-\I {\vec k} 
{\vec r}} \D \vec k,\; |k|=\sqrt{k_{x}^2 + k_{y}^2}. 
\end{equation} 
Using Hamiltonian (\ref{Hamiltonian}) and equations 
(\ref{Hamiltonian_equations}) one can get the dynamical equations 
\cite{Pushkarev1996}: 
\begin{equation} 
\label{eta_psi_equations} 
\begin{array}{rl} 
\displaystyle 
\dot \eta = &\hat k  \psi - (\nabla (\eta \nabla \psi)) - \hat k  [\eta \hat k  
\psi] +\\ 
\displaystyle 
&+ \hat k (\eta \hat k  [\eta \hat k  \psi]) + \frac{1}{2} \nabla^2 [\eta^2 \hat 
k \psi] +\\ 
\displaystyle 
&\frac{1}{2} \hat k [\eta^2 \nabla^2\psi] + \hat F^{-1}[\gamma_k \eta_k],\\ 
\displaystyle 
\dot \psi = &- g\eta - \frac{1}{2}\left[ (\nabla \psi)^2 - (\hat k \psi)^2 
\right] - \\ 
\displaystyle 
&- [\hat k  \psi] \hat k  [\eta \hat k  \psi] - [\eta \hat k  \psi]\nabla^2\psi  
+ \hat F^{-1}[\gamma_k \psi_k]. 
\end{array} 
\end{equation} 
Here $\hat F^{-1}$ corresponds to inverse Fourier transform. We introduced 
artificial dissipative terms 
$\hat F^{-1}[\gamma_k \psi_k]$, corresponding to pseudo-viscous high frequency 
damping.

It is important to stress that we added dissipation terms in {\underline both}
equations. In fact, equation for $\dot\eta$ is just kinematic boundary condition,
and adding a smoothing term to this equation has no any physical sense. Nevertheless,
adding of this term is necessary for stability of the numerical scheme.
 
The model (\ref{Hamiltonian_equations})-(\ref{eta_psi_equations}) was used in 
the numerical experiments \cite{Pushkarev1996} -- \cite{Pushkarev2000}, 
\cite{DKZ2003_Grav}, \cite{DKZ2004}, \cite{Mesoturb2005}, \cite{Nazarenko2005}, \cite{Nazarenko2006}. 
 
Introduction of the complex normal variables $a_{\vec k}$ 
\begin{equation} 
a_{\vec k} = \sqrt \frac{\omega_k}{2k} \eta_{\vec k} + \I \sqrt 
\frac{k}{2\omega_k} \psi_{\vec k}, 
\end{equation} 
where $\omega_k = \sqrt {gk}$, transforms equations 
(\ref{Hamiltonian_equations}) into 
\begin{equation} 
\frac{\partial a_{\vec k}}{\partial t} = -\I\frac{\delta H}{\delta a_{\vec 
k}^{*}}. 
\end{equation} 
 
To proceed with statistical description of the wave ensemble, first, one should 
perform the 
canonical transformation $a_{\vec k} \rightarrow b_{\vec k}$, which excludes the 
cubical terms 
in the Hamiltonian. The details of this transformation can be found in the paper
by Zakharov (1999) \cite{Zakharov1999}.
After the transformation the Hamiltonian takes the forms 
\begin{equation} 
\label{Hamiltonian_b_k} 
\begin{array}{c} 
\displaystyle 
H = \int \omega_{\vec k} b_{\vec k} b^{*}_{\vec k} + \frac{1}{4}\int T_{\vec k 
\vec k_1 \vec k_2 \vec k_3} 
b_{\vec k}^{*} b_{\vec k_1}^{*} b_{\vec k_2} b_{\vec k_3}\times\\ 
\displaystyle 
\times\delta_{\vec k + \vec k_1 - \vec k_2 - \vec k_3}\D\vec k_1 \D\vec k_2 \D\vec 
k_3. 
\end{array} 
\end{equation} 
where $T$ is a homogeneous function of the third order: 
\begin{equation} 
T(\varepsilon\vec k, \varepsilon\vec k_1, \varepsilon\vec k_2, \varepsilon\vec 
k_3) = 
\varepsilon^3 T(\vec k, \vec k_1, \vec k_2, \vec k_3). 
\end{equation} 
Connection between $a_{\vec k}$ and $b_{\vec k}$ together with explicit 
expression for 
$T_{\vec k \vec k_1 \vec k_2 \vec k_3}$ can be found, for example, in 
\cite{Zakharov1999}. 
 
Let us introduce the pair correlation function 
\begin{equation} 
<a_{\vec k} a_{\vec k'}^{*}> = g N_{\vec k} \delta(\vec k - \vec k'), 
\end{equation} 
where $N_{\vec k}$ is the spectral density of the wave function. This definition 
of the wave 
action is common in oceanography. 
 
We also introduce the correlation function for transformed normal variables 
\begin{equation} 
<b_{\vec k} b_{\vec k'}^{*}> = g n_{\vec k} \delta(\vec k - \vec k') 
\end{equation} 
Functions $n_{\vec k}$ and $N_{\vec k}$ can be expressed through each other in 
terms of cumbersome 
power series \cite{Zakharov1999}. On deep water their relative difference is of 
the order of $\mu^2$ 
($\mu$ is the characteristic steepness) and can be neglected (in most cases of swell 
evolution (or wave evolution) experimental results shows $\mu \simeq 0.1$). 
 
Spectrum $n_{\vec k}$ satisfies Hasselmann (kinetic) equation \cite{Hasselmann1962} 
\begin{equation} 
\label{Hasselmann_equation} 
\begin{array}{l} 
\displaystyle 
\frac{\partial n_{\vec{k}}}{\partial t}=S_{nl}[n] + S_{diss} + 2\gamma_k n_{\vec 
k}, \\ 
\displaystyle 
S_{nl}[n]=2\pi g^2 \int |T_{\vec{k},\vec{k_1},\vec{k_2},\vec{k_3}}|^2 
\left(n_{\vec{k_1}}n_{\vec{k_2}}n_{\vec{k_3}}+\right.\\ 
\displaystyle 
\left. + n_{\vec{k}}n_{\vec{k_2}}n_{\vec{k_3}} - 
n_{\vec{k}}n_{\vec{k_1}}n_{\vec{k_2}} - 
n_{\vec{k}}n_{\vec{k_1}}n_{\vec{k_3}}\right)\times\\ 
\displaystyle 
\times\delta\left( 
\omega_k+\omega_{k_1}-\omega_{k_2}-\omega_{k_3}\right)\times\\ 
\displaystyle 
\times\delta\left(\vec{k}+\vec{k_1}-\vec{k_2}-\vec{k_3}\right)\,\D 
\vec{k_1}\D\vec{k_2} \D\vec{k_3}. 
\end{array} 
\end{equation} 
Here $S_{diss}$ is an empiric dissipative term, corresponding to white-capping. 
 
Stationary conservative kinetic equation 
\begin{equation} 
S_{nl} = 0 
\end{equation} 
has the rich family of Kolmogorov-type \cite{Kolmogorov1941} exact solutions. 
Among them is Zakharov-Filonenko spectrum \cite{Zakharov1966} for the direct 
cascade of energy 
\begin{equation} 
n_k \sim \frac{1}{k^4}, 
\end{equation} 
and Zakharov-Zaslavsky \cite{ZakharovPhD}, \cite{Zakharov1982} spectra for the 
inverse cascade 
of wave action 
\begin{equation} 
n_k \sim \frac{1}{k^{23/6}}, 
\end{equation} 
 
\section{Deterministic Numerical Experiment.} 
 
\subsection{Problem Setup} 
 
The dynamical equations (\ref{eta_psi_equations}) have been solved in the 
real-space domain $2\pi \times 2\pi$ 
on the grid $512 \times 4096$ with the gravity acceleration set to $g=1$. The 
solution has been performed by the 
spectral code, developed in \cite{KorotkevichPhD} and previously used in 
\cite{DKZ2003_Cap},\cite{DKZ2003_Grav}, 
\cite{DKZ2004},\cite{Mesoturb2005}. We have to stress that in the current
computations the resolution in 
$Y$-direction (long axis) is better than the resolution in $X$-direction by the factor of 
$8$. 
 
\begin{figure}[!htb] 
\centering 
\includegraphics[width=3.0in]{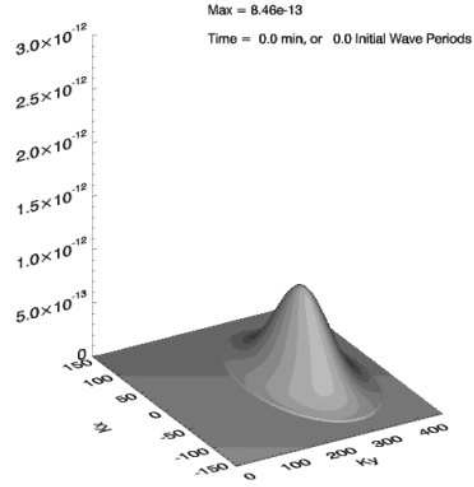} 
\caption{\label{InitialConditions3D}Initial distribution of $|a_{\vec k}|^2$ on 
$\vec k$-plane.} 
\end{figure}

This approach is reasonable if the swell is essentially anisotropic, almost 
one-dimensional. This assumption will be validated by the proper choice of 
the initial data for computation. As the initial condition, we used the  Gaussian-shaped distribution in Fourier 
space (see Fig.~\ref{InitialConditions3D}): 
\begin{equation} 
\label{Dynamic_initial_conditions} 
\begin{array}{l} 
\displaystyle 
\left\{ 
\begin{array}{l} 
\displaystyle 
|a_{\vec k}| = A_i \exp \left(- \frac{1}{2}\frac{\left|\vec k - \vec 
k_0\right|^2}{D_i^2}\right), 
\left|\vec k - \vec k_0\right| \le 2D_i,\\ 
\displaystyle 
|a_{\vec k}| = 10^{-12}, \left|\vec k - \vec k_0\right| > 2D_i, 
\end{array} 
\right.\\ 
\displaystyle 
A_i = 0.92\times10^{-6}, D_i = 60, \vec k_0 = (0; 300), \omega_0 = \sqrt{g k_0}. 
\end{array} 
\end{equation} 
The initial phases of all harmonics were random. The average steepness of this initial condition was $\mu \simeq 0.167$.

To realize similar experiment in the laboratory wave tank, one has 
to to generate the waves with wave-length $300$ times less than the length of 
the tank. The width of the tank would not be less than $1/8$  of its length. The 
minimal wave length of the gravitational wave in absence of capillary effects 
can be estimated as $\lambda_{min}\simeq 3 cm$. The leading wavelength should be 
higher by the order of magnitude $\lambda\simeq 30 cm$. 
 
In such big tank of $200 \times 25 $ meters experimentators can observe the 
evolution of the swell until approximately $700 T_0$ -- still less than in our 
experiments. In the tanks of smaller size, the effects of discreetness the 
Fourier space will be dominating, and experimentalists will observe either 
``frozen'', or ``mesoscopic'' wave turbulence, qualitatively different from the 
wave turbulence in the ocean.

To stabilize high-frequency numerical instability, the damping function has been 
chosen as 
\begin{equation} 
\label{Pseudo_Viscous_Damping} 
\begin{array}{l} 
\displaystyle 
\gamma_k = \left\{ 
\begin{array}{l} 
\displaystyle 
0, k < k_d,\\ 
\displaystyle 
- \gamma (k - k_d)^2, k \ge k_d,\\ 
\end{array} 
\right.\\ 
\displaystyle 
k_d = 1024, \gamma = 5.65 \times 10^{-3}. 
\end{array} 
\end{equation} 
 
The simulation was performed until $t = 336$, which is equivalent to $926 T_0$, 
where $T_0$ is the 
period of the wave, corresponding to the maximum of the initial spectral 
distribution. 
 
\subsection{Zakharov-Filonenko spectra} 
 
\begin{figure}[!htb] 
\centering 
\includegraphics[width=8.5cm,angle=0]{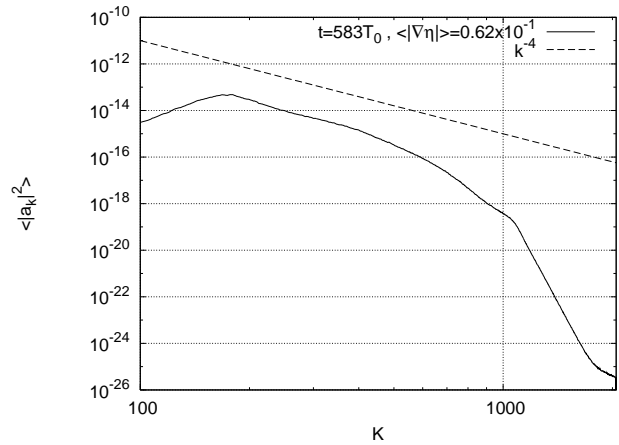} 
\caption{\label{Kolmogorov_k}Angle-averaged spectrum $n_k=<|a_{\vec k}|^2>$ in a 
double logarithmic scale. The tail of distribution 
fits to Zakharov-Filonenko spectrum.} 
\end{figure}

Like in the previous papers 
\cite{Onorato2002},\cite{DKZ2003_Grav},\cite{DKZ2004} and \cite{Mesoturb2005}, 
we observed fast formation of the spectral tail, described by Zakharov-Filonenko 
law for the direct cascade 
$n_k \sim k^{-4}$ \cite{Zakharov1966} (see Fig.\ref{Kolmogorov_k}). In the 
agreement with \cite{Mesoturb2005}, 
the spectral maximum slowly down-shifts to the large scales region, which 
corresponds to the inverse cascade 
\cite{ZakharovPhD},\cite{Zakharov1982}. 
 
\begin{figure}[!htb] 
\centering 
\includegraphics[width=8.5cm,angle=0]{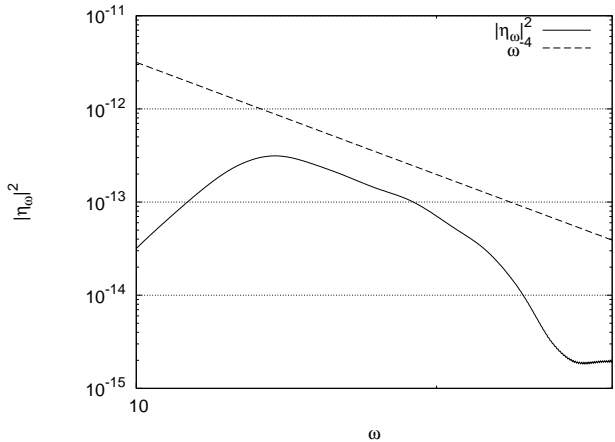} 
\caption{\label{Kolmogorov_omega}Energy spectrum in a double logarithmic scale. 
The tail of distribution fits to asymptotics $\omega^{-4}$.} 
\end{figure}

Also, the direct measurement of energy spectrum has been performed during the 
final stage of the simulation, 
when the spectral down shift was slow enough.  This experiment can be 
interpreted as the ocean buoy record -- 
the time series of the surface elevations has been recorded at one point of the 
surface during 
$T_{buoy} \simeq 300 T_0$. The Fourier transform of the autocorrelation function 
\begin{equation} 
E(\omega) = \frac{1}{2\pi}\int\limits_{-T_{buoy}/2}^{T_{buoy}/2}<\eta(t+\tau)\eta(\tau)>e^{i\omega t}\D \tau \D t. 
\end{equation} 
allows to detect the direct cascade spectrum tail proportional to $\omega^{-4}$ 
(see Fig.\ref{Kolmogorov_omega}), 
well known from experimental observations 
\cite{Toba1973},\cite{Donelan1985},\cite{Hwang2000}. 
 
\subsection{Is the weak-turbulent scenario realized?} 
 
Presence of Kolmogorov asymptotics in spectral tails, however, is not enough to 
validate applicability of the 
weak-turbulent scenario for description of wave ensemble. We have also to be sure 
that statistical properties of 
this ensemble correspond to weak-turbulent theory assumptions. 
 
\begin{figure}[!htb] 
\centering 
\includegraphics[width=3.0in]{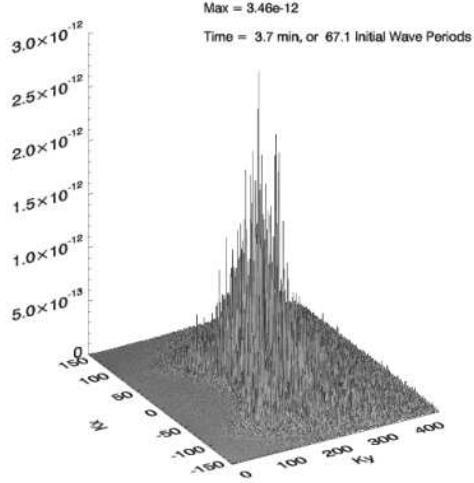}
\caption{\label{RawDistribution}Surface $|a_{\vec k}|^2$ at the moment of time 
$t\simeq67T_0$.} 
\end{figure}

We should stress that in our experiments at thebeginning $|a_{\vec k}|^2$ is a smooth function
of $\vec k$. Only phases of individual waves are random.
As shows numerical simulation, the initial condition 
(\ref{Dynamic_initial_conditions}) (see Fig.\ref{InitialConditions3D}) 
does not preserve its smoothness -- it becomes rough almost immediately (see 
Fig.\ref{RawDistribution}). 
The picture of this roughness is remarkably preserved in many details, even as 
the spectrum down-shifts as a whole. This roughness does not contradict the 
weak-turbulent theory. According to this theory, the wave ensemble is almost 
Gaussian, and both real and imaginary parts of each separate harmonics are 
not-correlated. However, according to the weak-turbulent theory, the spectra 
must become smooth after averaging over long enough time of more than $1/\mu^2$ 
periods.  Earlier we observed such restoring of smoothness in the numerical 
experiments of the $MMT$ model (see \cite{ZGDP},\cite{DPZ}, \cite{DGPZ} and \cite{GZD}). However, 
in the experiments discussed in the article, the roughness still persists and the 
averaging does not suppresses it completely. It can be explained by sparsity of 
the resonances. 
 
Resonant conditions are defined by the system of equations: 
\begin{equation} 
\label{resonant_conditions} 
\begin{array}{c} 
\displaystyle 
\omega_k + \omega_{k_1} = \omega_{k_2} + \omega_{k_3},\\ 
\displaystyle 
\vec k + \vec k_1 = \vec k_2 + \vec k_3, 
\end{array} 
\end{equation} 
These resonant conditions define five-dimensional hyper-surface in 
six-dimensional space $\vec k, \vec k_1, \vec k_2$. In any finite system, 
(\ref{resonant_conditions}) turns into Diophantine equation. Some solutions of 
this equation are known \cite{DZ1994}, 
\cite{Nazarenko2005}. In reality, however, the energy transport is realized not 
by exact, but "approximate" resonances,
posed in a layer near the resonant surface and defined by 
\begin{equation} 
\label{resonant_layer} 
|\omega_{k} + \omega_{k_1} - \omega_{k_2} - \omega_{k + k_1 - k_2}| \le 
\Gamma, 
\end{equation} 
where $\Gamma$ is a characteristic inverse time of nonlinear interaction. 
 
In the finite systems $\vec k, \vec k_1, \vec k_2$ take values on the nodes of 
the discrete grid. The weak turbulent approach is valid, 
if the density of discrete approximate resonances inside the layer 
(\ref{resonant_layer}) is high enough. In our case the lattice 
constant is $\Delta k = 1$, and typical relative deviation from the resonance 
surface 
\begin{equation} 
\frac{\Delta \omega}{\omega} \simeq \frac{\omega_k'}{\omega} \Delta k = 
\frac{\omega_k'}{\omega} \simeq \frac{1}{600} 
\simeq 2\times10^{-3}. 
\end{equation} 
Inverse time of the interaction $\Gamma$ can be estimated from our numerical 
experiments: wave amplitudes change essentially during 30 
periods, and one can assume: $\Gamma/\omega \simeq 10^{-2} \gg \frac{\delta 
\omega}{\omega}$. It means that the condition for the 
applicability of weak turbulent theory is typically satisfied, but the "reserve" 
for their validity is rather modest. As a result, 
some particular harmonics, posed in certain "privileged" point of $k$-plane 
could form a "network" of almost resonant quadruplets 
and realize significant part of energy transport. Amplitudes of these harmonics 
exceed the average level essentially. This effect 
was described in the article \cite{Mesoturb2005}, where such "selected few" 
harmonics were called "oligarchs". If "oligarchs" 
realize most part of the energy flux, the turbulence is "mesoscopic", not weak. 
 
\subsection{Statistics of the harmonics} 
 
According to the weak-turbulent scenario, statistics of the $a_{\vec k}(t)$ in 
any given $k$ should be close to 
Gaussian. It presumes that the $PDF$ for the squared amplitudes is 
\begin{equation}
\label{GaussPDF}
P(|a_{\vec k}|^2) \simeq \frac{1}{D}e^{-|a_{\vec k}|^2/D}, 
\end{equation} 
here $D = <|a_{\vec k}|^2>$ --- mean square amplitude.
To check equation (\ref{GaussPDF}) we need to find a way for calculation of $D(\vec k)$.
If the ensemble is 
stationary in time, $D(\vec k)$ could 
be found for any given $k$ by averaging in time. In our case, the process is 
non-stationary, and we have a problem 
with determination of $D(\vec k)$. 
 
\begin{figure}[!htb] 
\centering 
\includegraphics[width=3.0in]{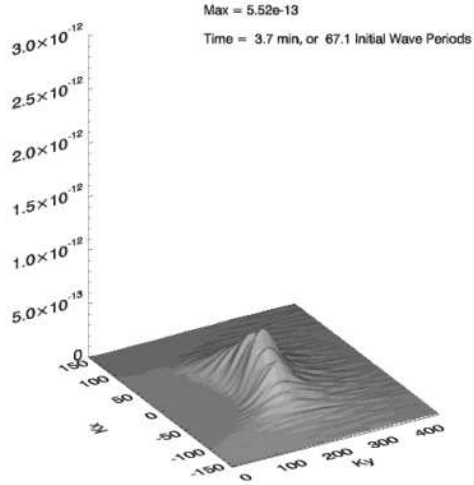} 
\caption{\label{SmoothedDistribution}Low-pass filtered surface $|a_{\vec k}|^2$ 
at $t\simeq 67 T_0$.} 
\end{figure}

\begin{figure}[!htb] 
\centering 
\includegraphics[width=8.5cm]{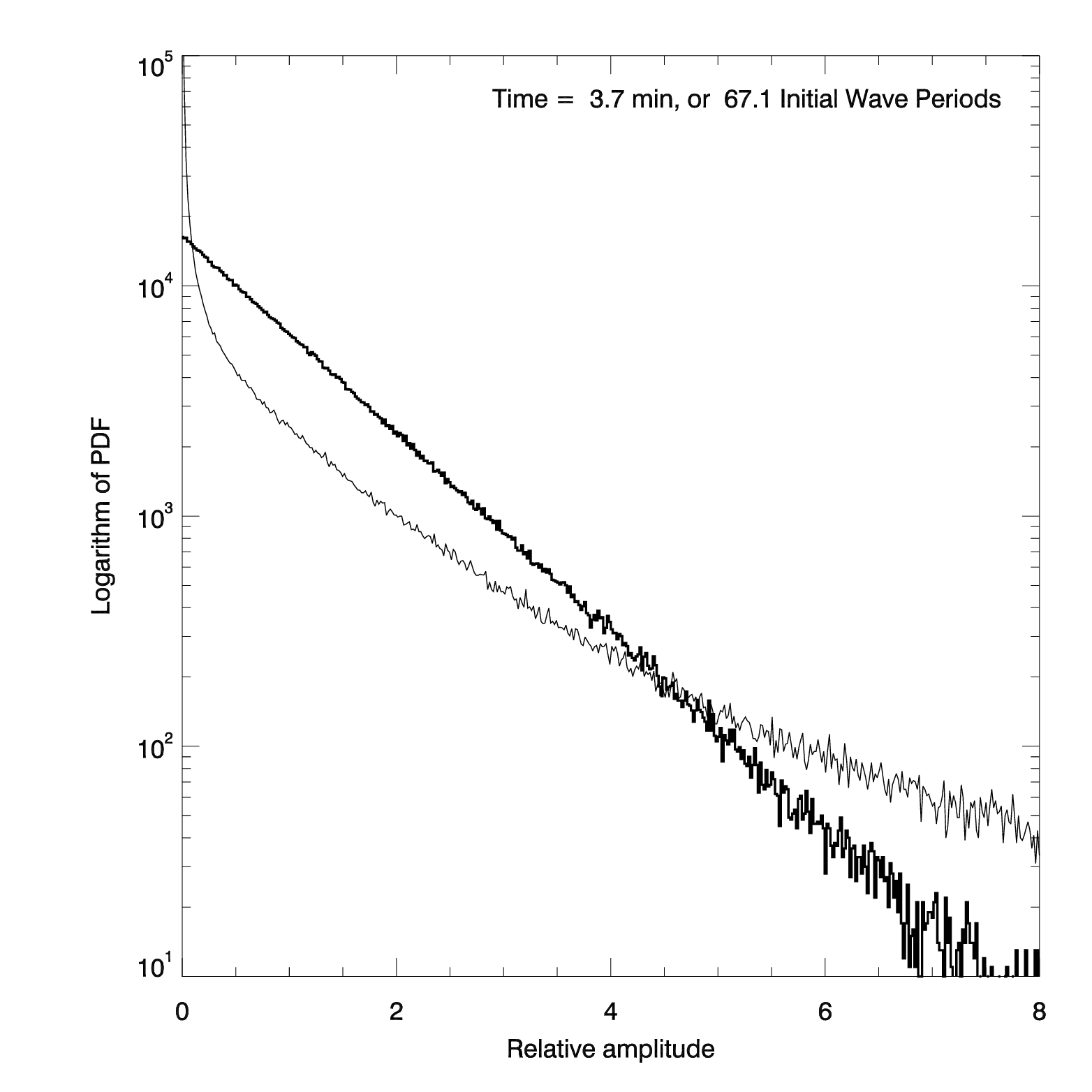} 
\caption{\label{DistributionFunction70}Probability distribution function (PDF) 
for relative squared amplitudes $|a_k|^2/<|a_k|^2>$. $t\simeq67T_0$.} 
\end{figure}

\begin{figure}[!htb] 
\centering 
\includegraphics[width=8.5cm]{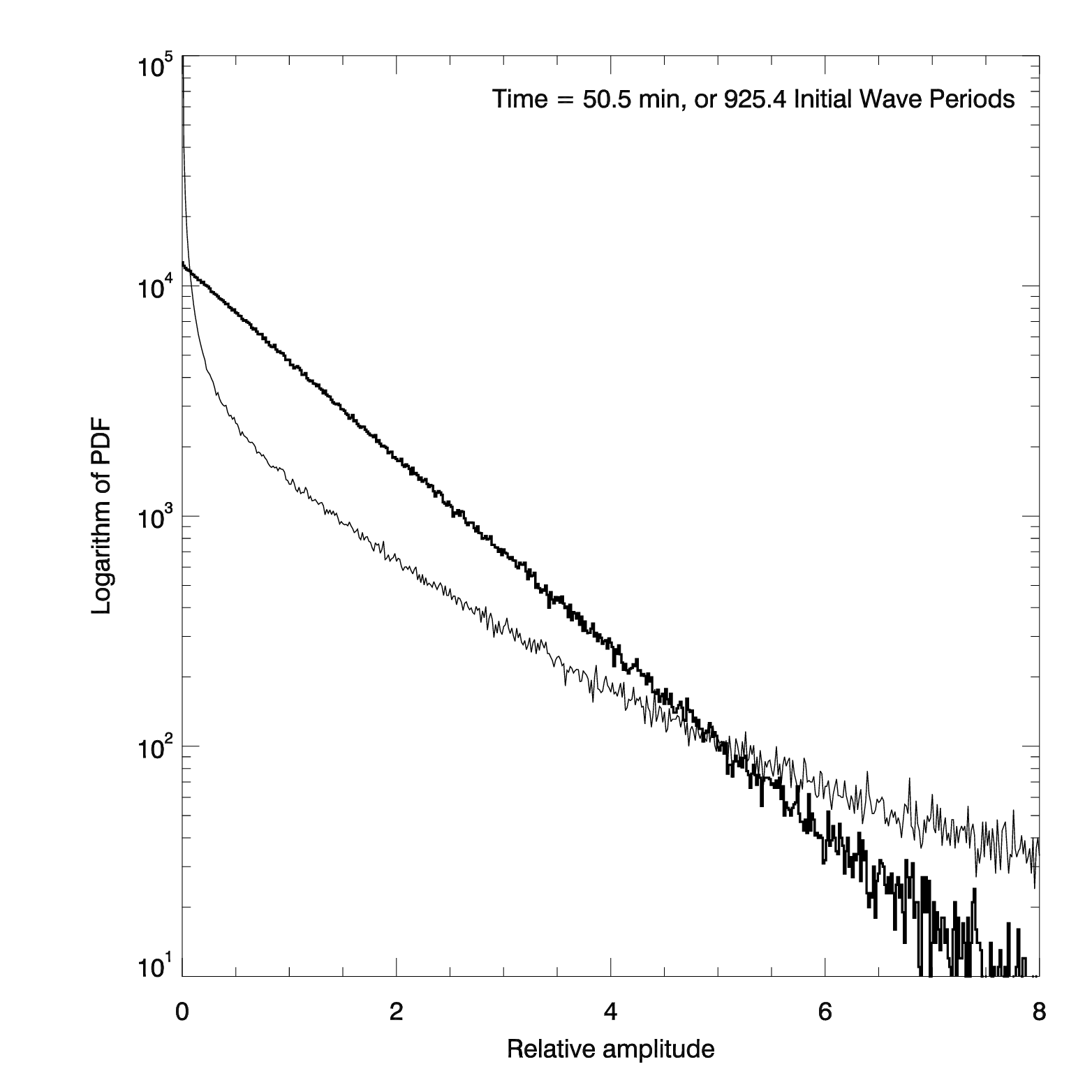} 
\caption{\label{DistributionFunction933}Probability distribution function (PDF) 
for relative squared amplitudes $|a_k|^2/<|a_k|^2>$. $t\simeq925T_0$.} 
\end{figure}

To resolve this problem, we used low-pass filtering instead of time averaging. 
The low-pass filter was chosen in the form 
\begin{equation} 
\begin{array}{l} 
\displaystyle 
f(\vec n) = e^{-(|\vec n|/\Delta)^3}, \Delta=0.25 Nx/2, Nx=4096. 
\end{array} 
\end{equation} 
This choice of the low-pass filter preserves the  values of total energy, wave 
action and the total momentum within three percent accuracy, see 
Fig.\ref{SmoothedDistribution}. 
Then it is possible to average the $PDF$ over different areas in $k$-space. The 
results for two different moments of time $t\simeq70T_0$ and $t\simeq933T_0$ 
are presented in Fig.\ref{DistributionFunction70} and 
Fig.\ref{DistributionFunction933}. 
The thin line gives $PDF$ after averaging over dissipation region harmonics, 
while bold line presents averaging over the 
non-dissipative area $|\vec k| < k_d = 1024$. One can see that statistics in the 
last case is quite close to the Gaussian, 
while in the dissipation region it deviates from Gaussian. However, deviation 
from the Guassianity in the dissipation region 
doesn't create any problems, since the "dissipative" harmonics do not contain 
any essential amount of the total energy, 
wave action and momentum. 
 
One should remember, that the bold lines in the Fig.\ref{DistributionFunction70} 
and Fig.\ref{DistributionFunction933} are 
the results of averaging over a million of harmonics. Among them there is a 
population of "selected few", or "oligarchs", 
with the amplitudes exceeding the average value by the factor of more than ten 
times. The "oligarchs" exist because our 
grid is still not fine enough. 
 
In our case "oligarchs" do exist, but their contribution in the total wave 
action is not more 4\%. Ten leading "oligarchs" at the end 
of the experiment are presented in the Appendix A.

\subsection{Two-stage evolution of the swell} 
 
\begin{figure}[ht]
\centering
\includegraphics[scale=0.43]{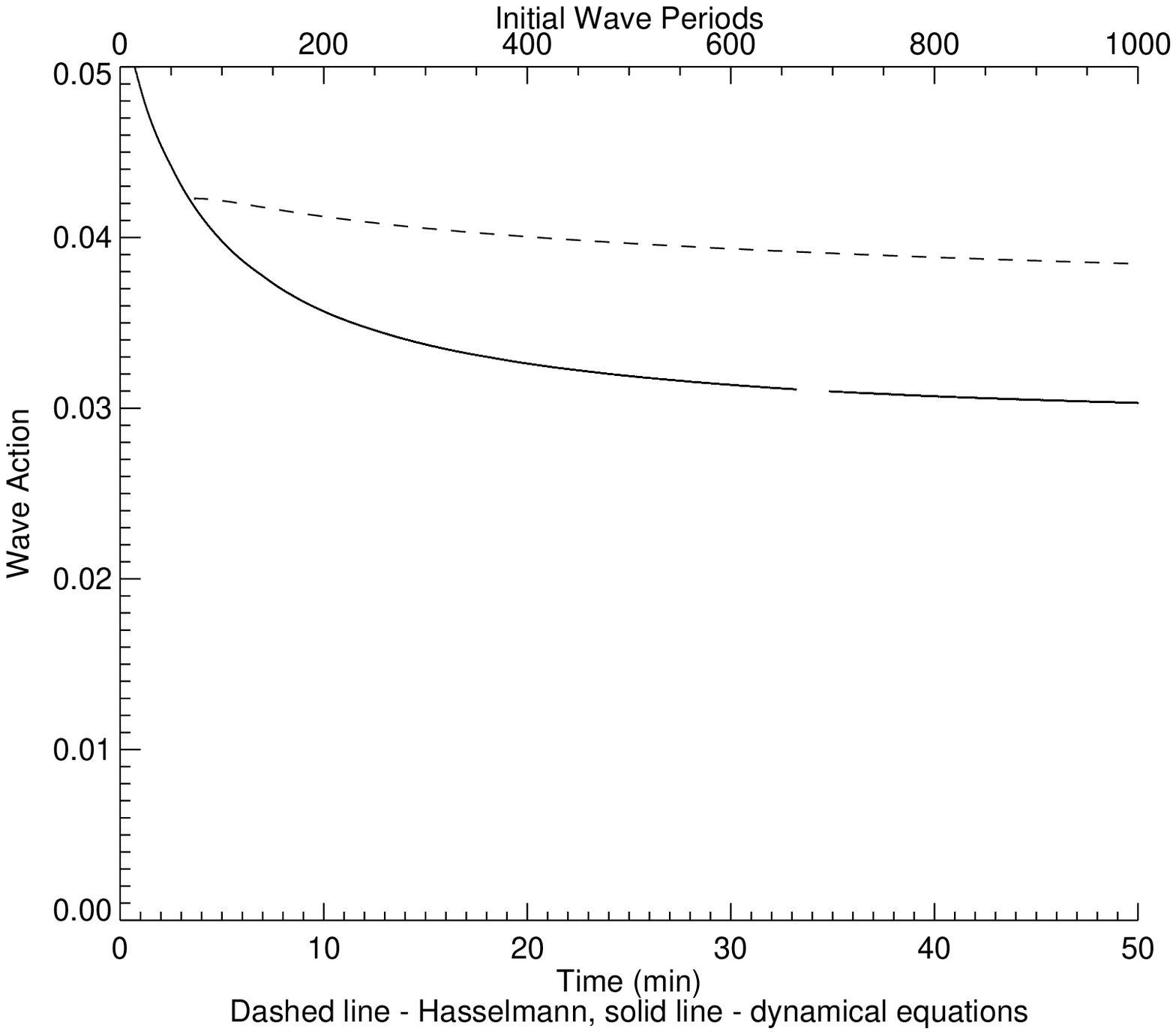} 
\caption{Total wave action as a function of time for the artificial viscosity 
case.}\label{N_ArtVisc} 
\end{figure}

\begin{figure}[ht]
\centering 
\includegraphics[scale=0.43]{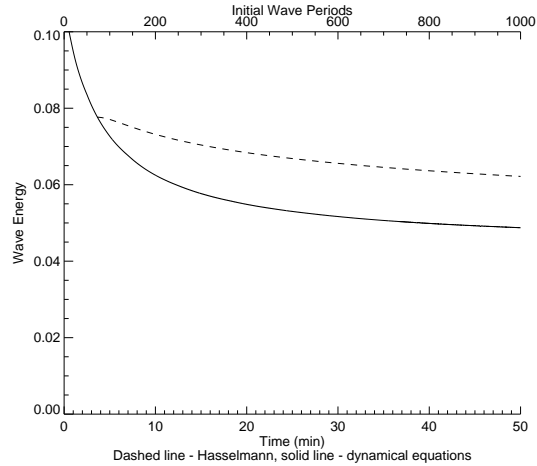} 
\caption{Total wave energy as a function of time for the artificial viscosity 
case}\label{H_ArtVisc} 
\end{figure}

\begin{figure}[ht]
\centering 
\includegraphics[scale=0.43]{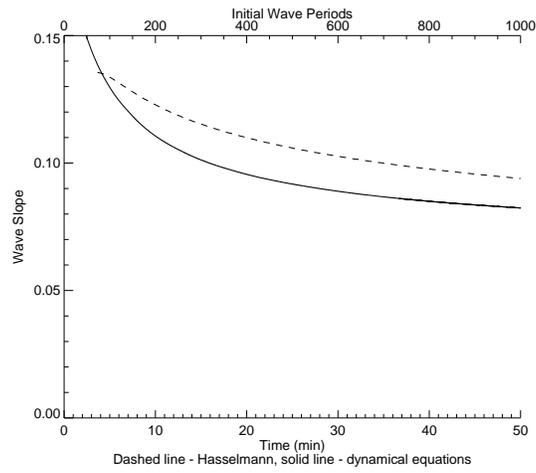} 
\caption{Average wave slope as a function of time for the artificial viscosity 
case.}\label{Slope_ArtVisc} 
\end{figure}

\begin{figure}[ht]
\centering 
\includegraphics[scale=0.43]{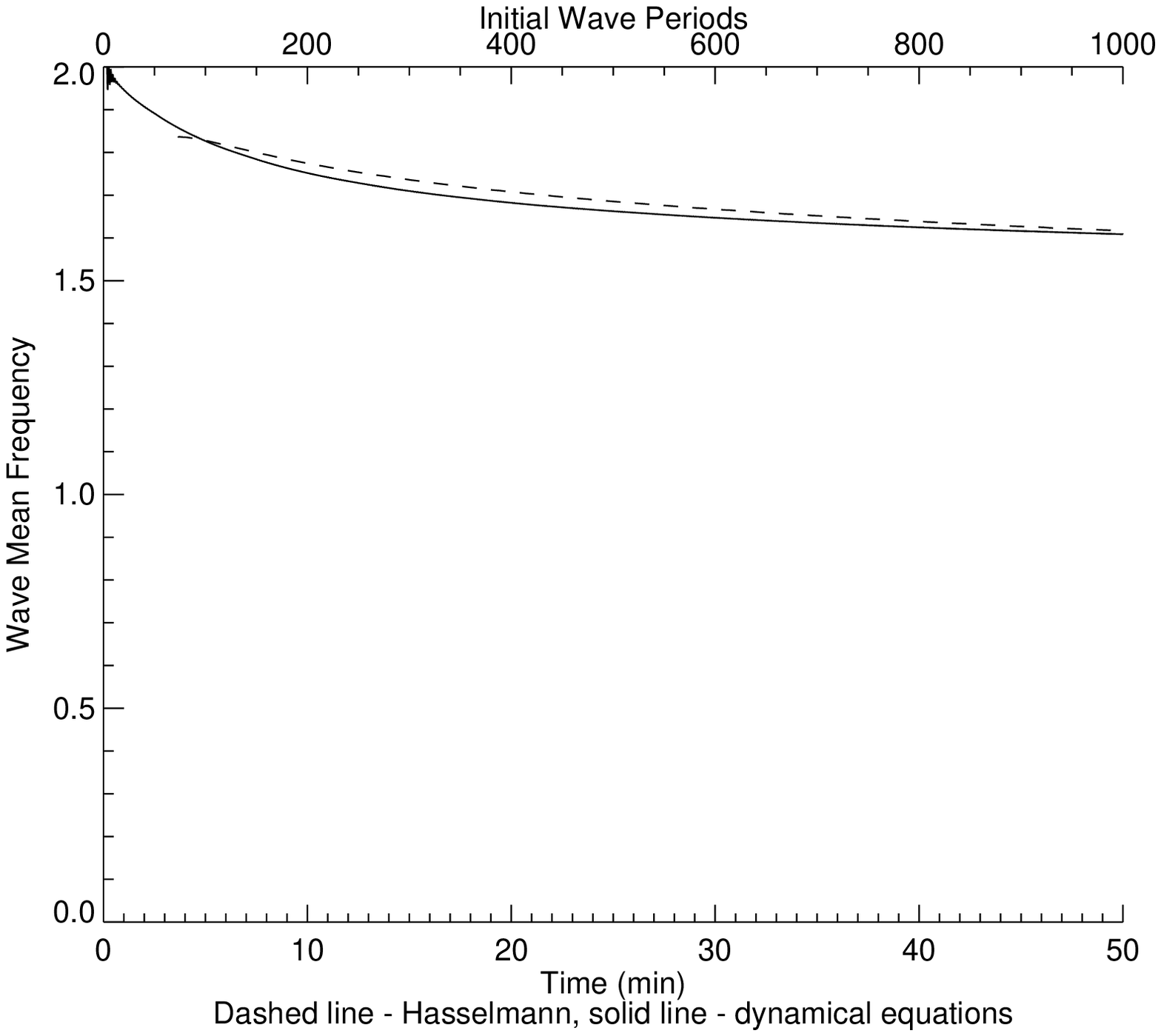} 
\caption{Mean wave frequency as a function of time for the artificial viscosity 
case.}\label{MeanFreq_ArtVisc} 
\end{figure}

Fig. \ref{N_ArtVisc}-\ref{MeanFreq_ArtVisc} demonstrate time evolution of 
main characteristics of the wave field: 
wave action, energy, characteristic slope and mean frequency. 

Fig.\ref{Slope_ArtVisc} should be specially commented. Here 
and further we define the characteristic slope as follows 
\begin{equation} 
\mu = \sqrt{2} \left[<(\nabla \eta)^2>\right]^{1/2}. 
\end{equation} 
Following this definition for the Stokes wave of small amplitude 
\begin{eqnarray*} 
\eta = a\cos(kx),\\ 
\mu = ak. 
\end{eqnarray*} 
According to this definition of steepness for the classical Pierson-Moscowitz 
spectrum $\mu = 0.095$. Our initial 
steepness $\mu\simeq0.167$ exceeds this value essentially. 
 
Evolution of the spectrum can be conventionally separated in two phases. On the 
first stage we observe fast 
drop of wave action, slope and especially energy. Then the drop is stabilized, 
and we observe slow down-shift 
of mean frequency together with angular spreading. Level lines of smoothed 
spectra in the first and in the 
last moments of time are shown in 
Fig.\ref{Spectrum_initial}-\ref{Spectrum_final} 
\begin{figure}[!htb] 
\centering 
\includegraphics[width=3.0in]{figures/bin.colour.a_k_c.000000.eps2} 
\caption{\label{Spectrum_initial}Initial spectrum $|a_{\vec k}|^2$. $t=0$.} 
\end{figure} 
\begin{figure}[!htb] 
\centering 
\includegraphics[width=3.0in]{figures/bin.colour.a_k_c.810000.eps2} 
\caption{\label{Spectrum_final}Final spectrum $|a_{\vec k}|^2$. 
$t\simeq933T_0$.}
\end{figure}
 
\begin{figure}[!htb] 
\centering 
\includegraphics[width=8.5cm,angle=0]{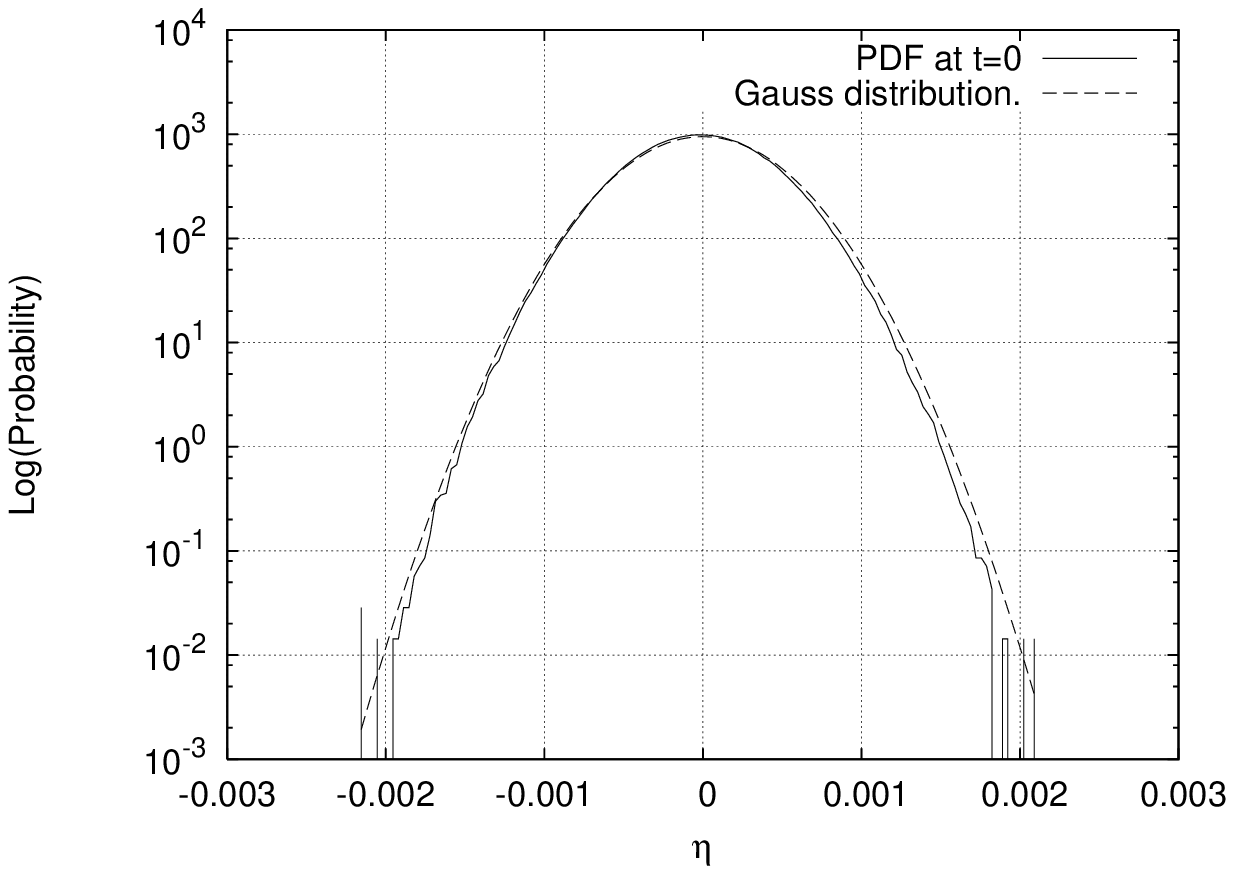} 
\caption{\label{PDF_eta_initial}PDF for the surface elevation $\eta$ at the 
initial moment of time. $t=0$.} 
\end{figure}

\begin{figure}[!htb] 
\centering 
\includegraphics[width=8.5cm,angle=0]{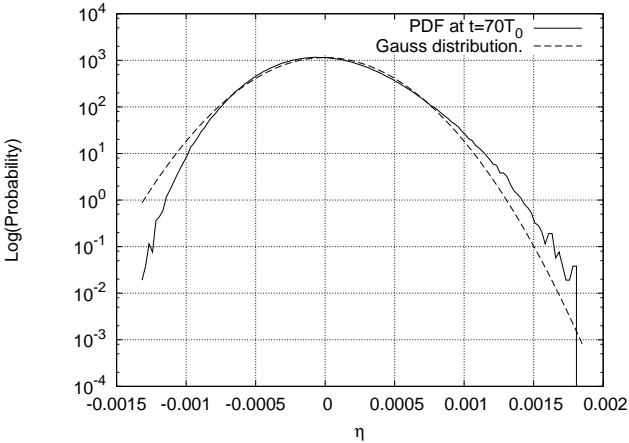} 
\caption{\label{PDF_eta_middle}PDF for the surface elevation $\eta$ at some 
middle moment of time. $t\simeq70T_0$.} 
\end{figure}

\begin{figure}[!htb] 
\centering 
\includegraphics[width=8.5cm,angle=0]{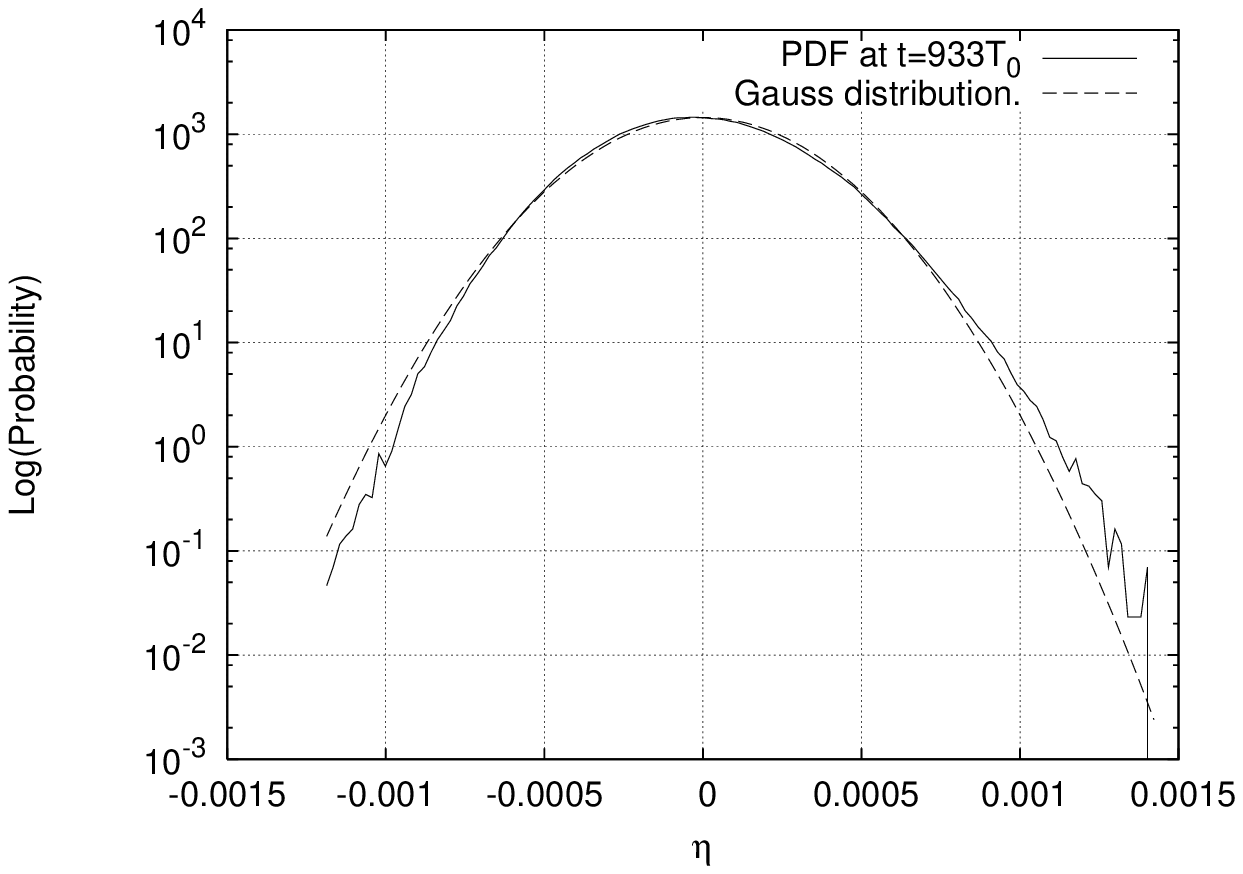} 
\caption{\label{PDF_eta_final}PDF for the surface elevation $\eta$ at the final 
moment of time. $t\simeq933T_0$.} 
\end{figure}

\begin{figure}[!htb] 
\centering 
\includegraphics[width=8.5cm,angle=0]{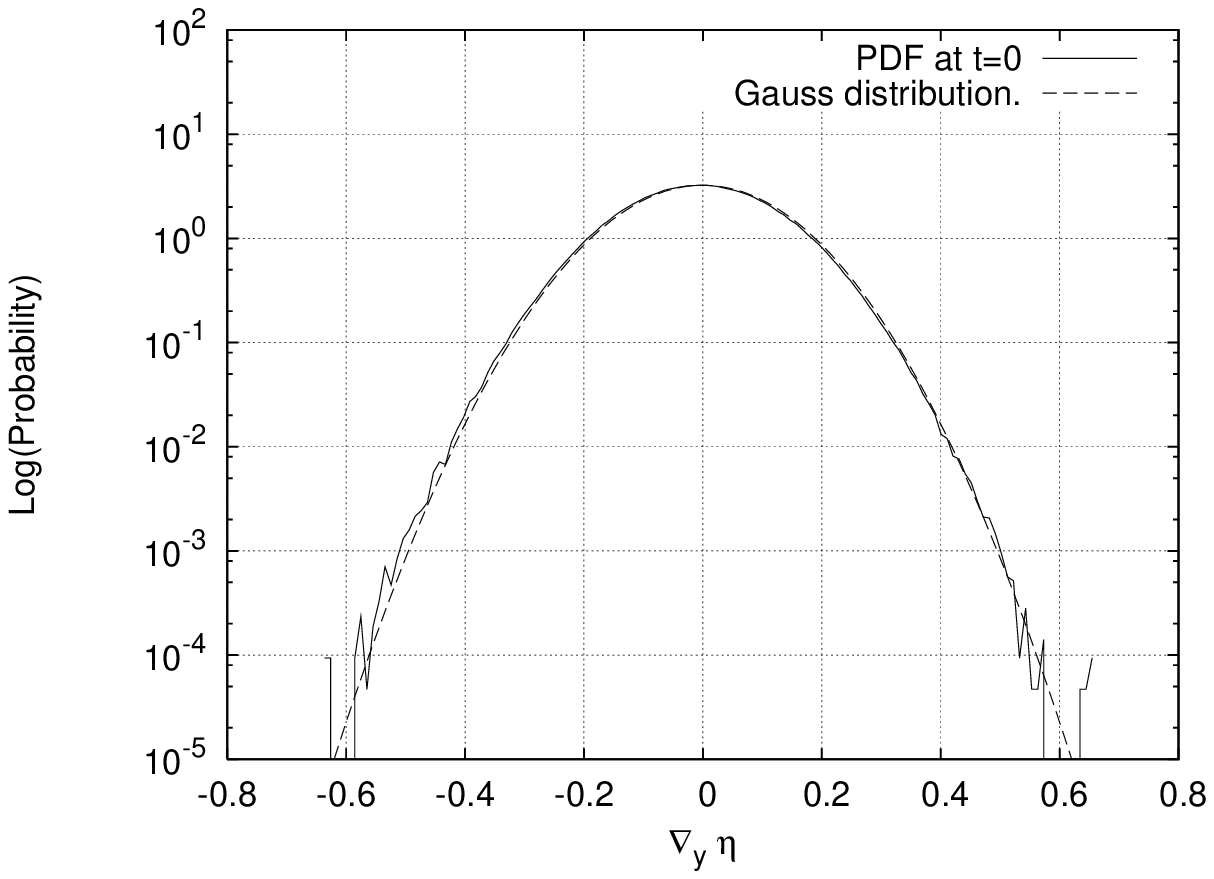} 
\caption{\label{GradY_initial}PDF for $(\nabla \eta)_y$ at the initial moment of 
time. $t=0$.} 
\end{figure}

\begin{figure}[!htb] 
\centering 
\includegraphics[width=8.5cm]{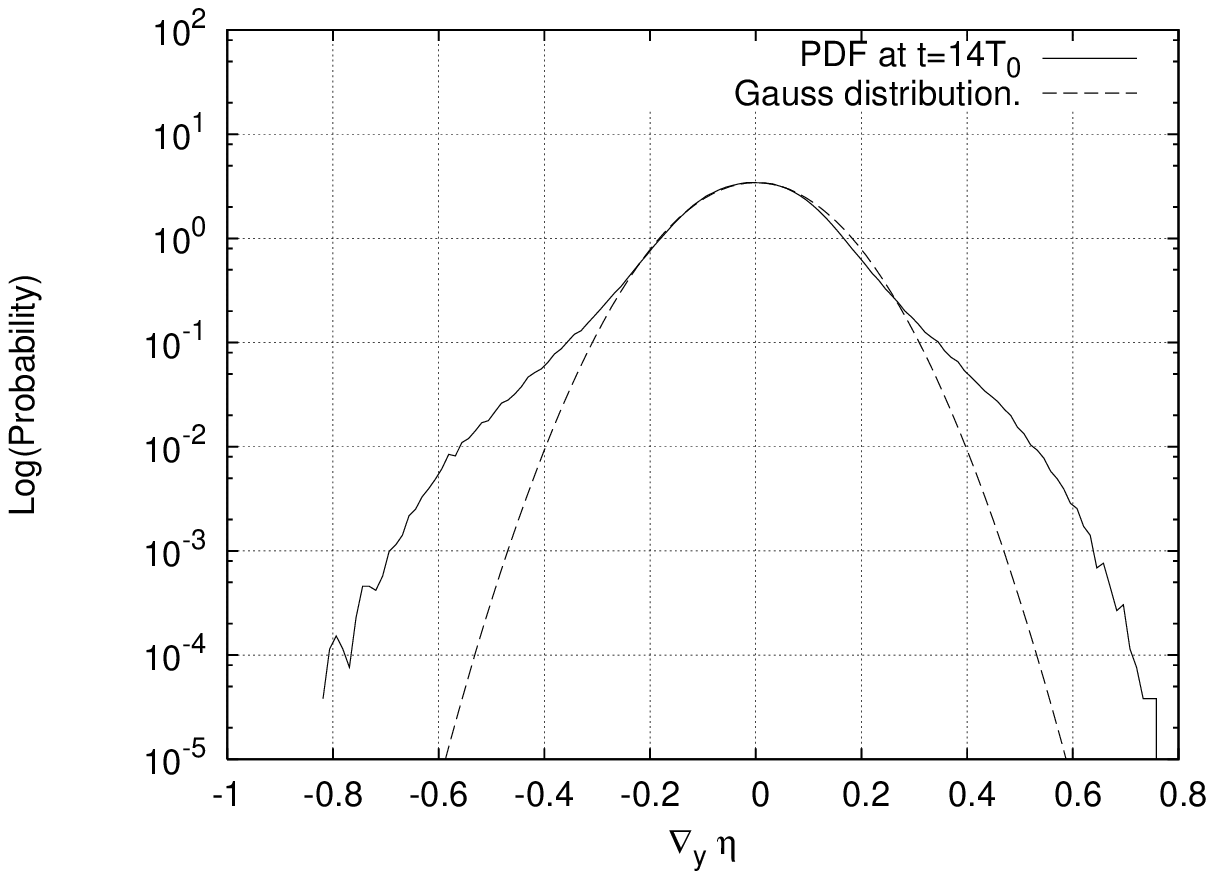} 
\caption{\label{GradY_max}PDF for $(\nabla \eta)_y$ at some middle moment of 
time. $t\simeq14T_0$.} 
\end{figure}

\begin{figure}[!htb] 
\centering 
\includegraphics[width=8.5cm,angle=0]{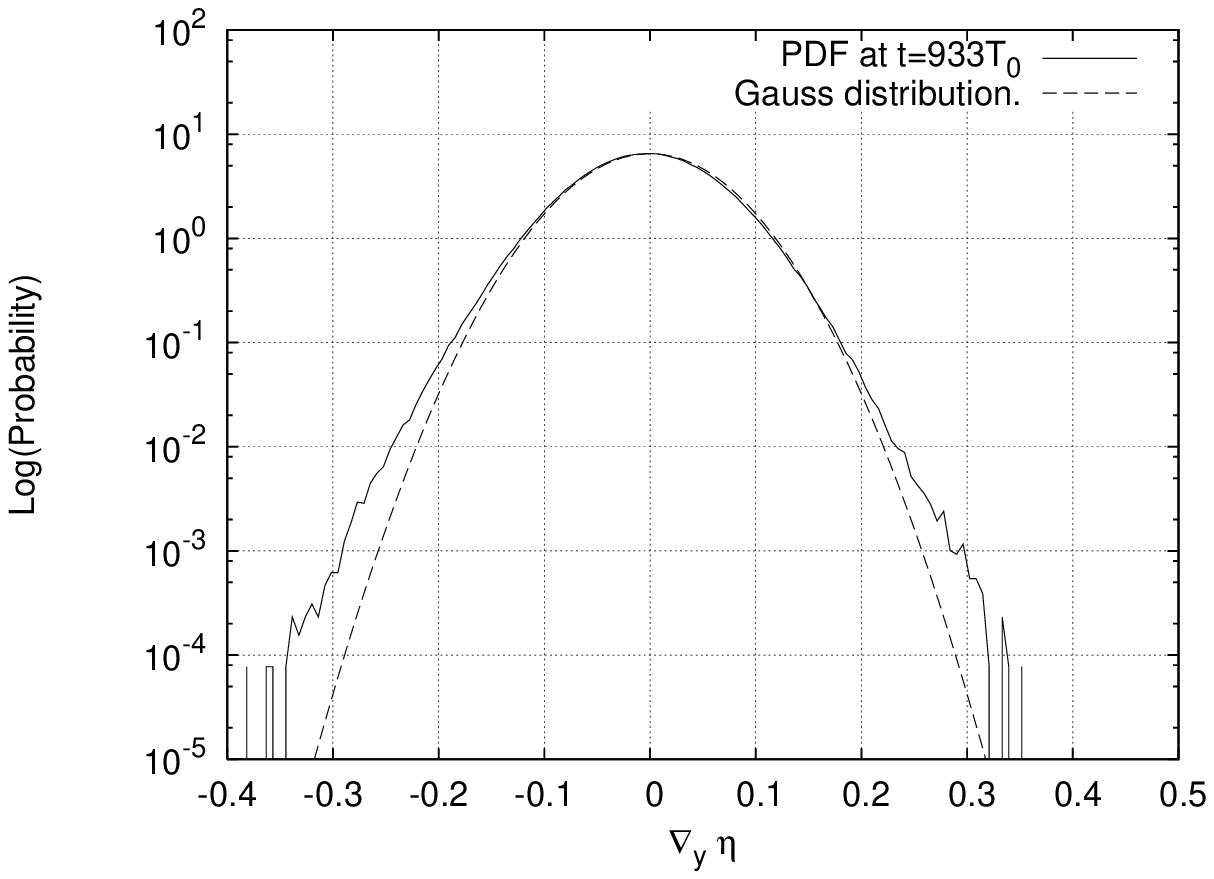} 
\caption{\label{GradY_final}PDF for $(\nabla \eta)_y$ at the final moment of 
time. $t\simeq933T_0$.} 
\end{figure}

Presence of two stages can be explained by study of the PDFs for elevation of 
the surface. In the initial 
moment of time PDF is Gaussian (Fig.\ref{PDF_eta_initial}). However, very soon 
intensive super-Gaussian tails appear 
(Fig.\ref{PDF_eta_middle}). Then they 
decrease slowly, and in the last moment of simulation, when characteristics of 
the sea are close to 
Peirson-Moscowitz, statistic is close to Gaussian again 
(Fig.\ref{PDF_eta_final}). Moderate tails do exist and, 
what is interesting, the PDF is not symmetric --- elevations are more probable 
troughs. 
PDF for $\eta_y$ --- longitudinal gradients in the first moments of time 
is Gaussian (Fig.\ref{GradY_initial}). Then in a very short period of time 
strong non-Gaussian tails appear and reach their 
maximum at $t\simeq14T_0$ (Fig.\ref{GradY_max}). Here $T_0 = 2\pi/\sqrt{k_0}$ 
--- period of initial leading wave. Since this 
moment the non-Gaussian tails decrease. In the last moment of simulation they 
are essentially reduced(Fig.\ref{GradY_final}).

Fast growing of non-Gaussian tails can be explained by fast formation of coherent
harmonics. Indeed, $14T_0 \simeq 2\pi/(\omega_0 \mu)$ is a characteristic time of
nonlinear processes due to quadratic nonlinearity. Note that the fourth harmonic
in our system is fast decaying, Hence we cannot see "real" white caps.
 
\begin{figure}[!htb] 
\centering 
\includegraphics[width=8.5cm]{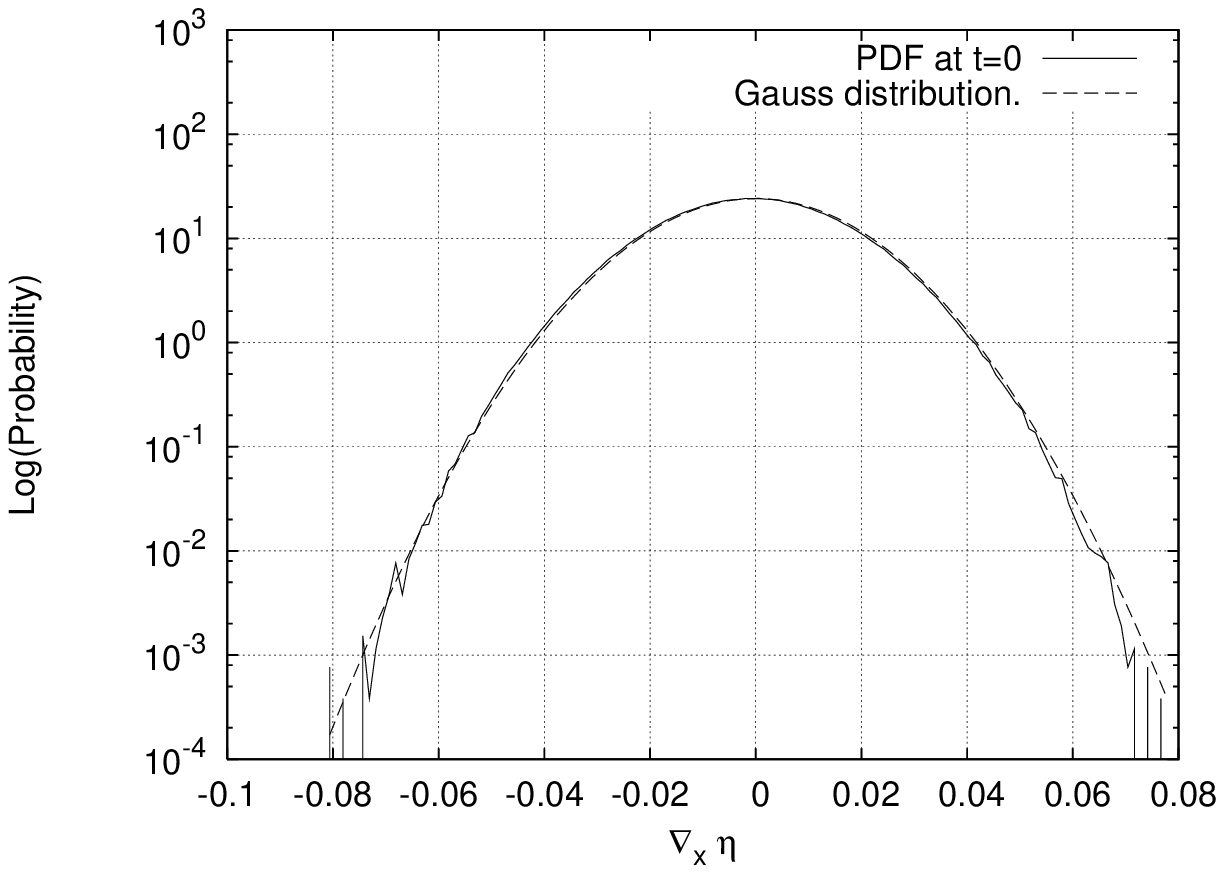} 
\caption{\label{GradX_initial}PDF for $(\nabla \eta)_x$ at the initial moment of 
time. $t=0$.} 
\end{figure}

\begin{figure}[!htb] 
\centering 
\includegraphics[width=8.5cm]{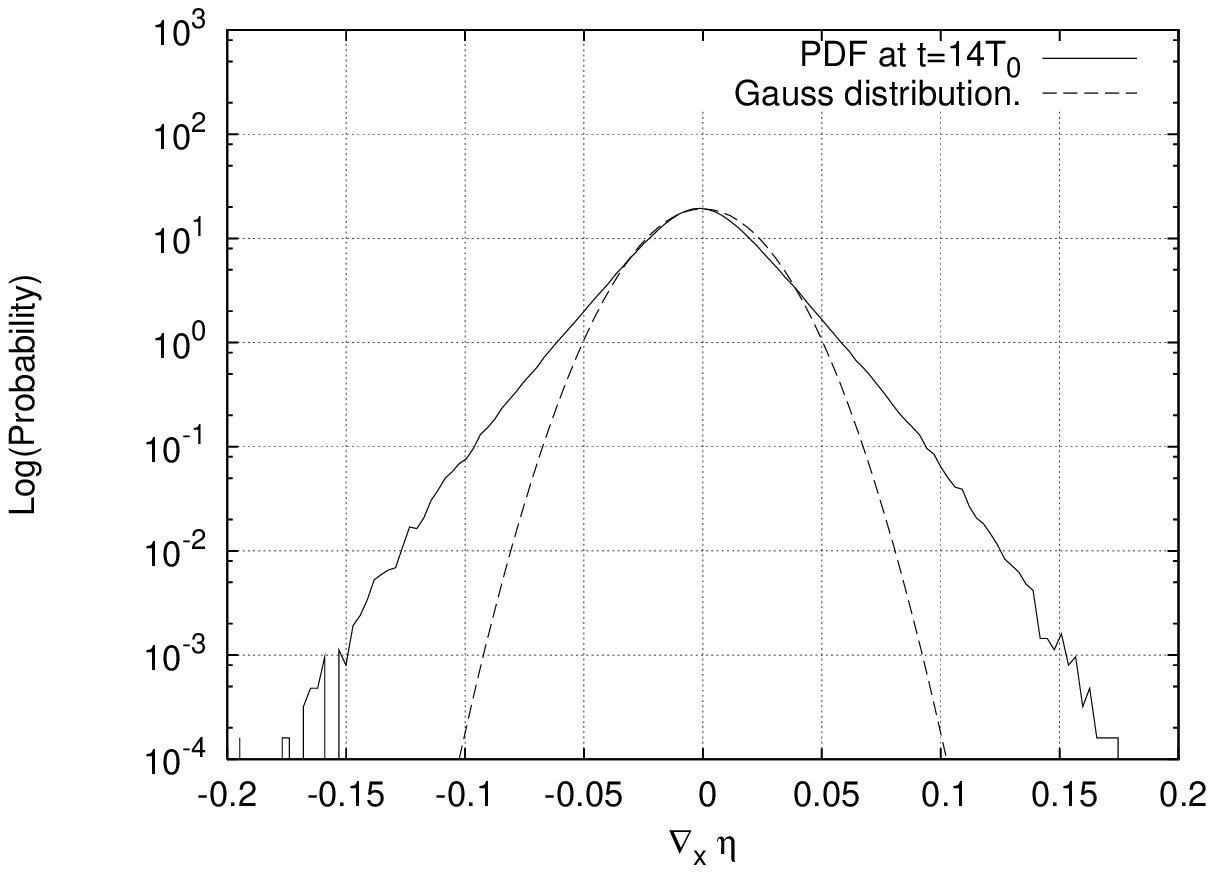} 
\caption{\label{GradX_max}PDF for $(\nabla \eta)_x$ at some middle moment of 
time. $t\simeq14T_0$.} 
\end{figure}

\begin{figure}[!htb] 
\centering 
\includegraphics[width=8.5cm]{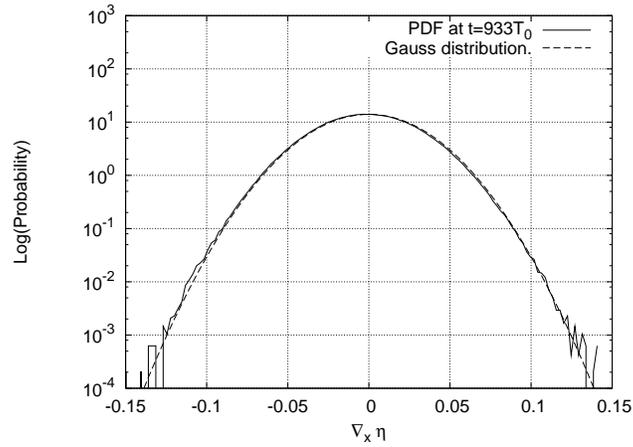} 
\caption{\label{GradX_final}PDF for $(\nabla \eta)_x$ at the final moment of 
time. $t\simeq933T_0$.} 
\end{figure}

\begin{figure}[!htb] 
\centering 
\includegraphics[width=3.0in]{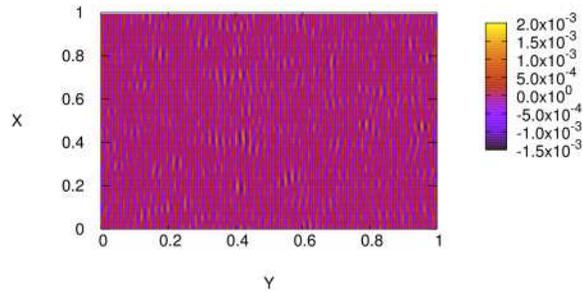} 
\caption{\label{Eta_surf_initial} Surface elevation at the initial moment of 
time. $t=0$.} 
\end{figure}

\begin{figure}[!htb] 
\centering 
\includegraphics[width=3.0in]{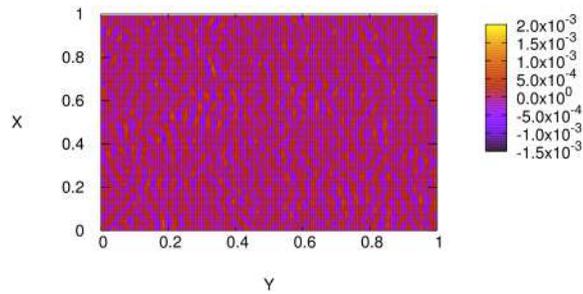} 
\caption{\label{Eta_surf_final} Surface elevation at the final moment of time. 
$t\simeq933T_0$.} 
\end{figure}

\begin{figure}[!htb] 
\centering 
\includegraphics[width=3.0in]{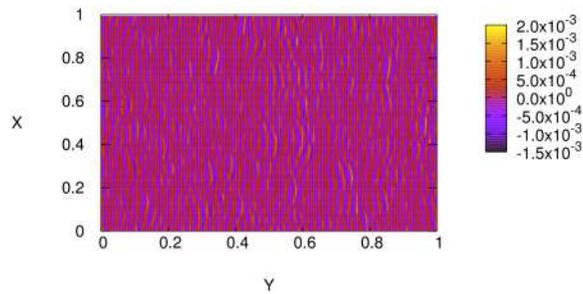} 
\caption{\label{Eta_surf_max} Surface elevation at the moment of maximum 
roughness. $t\simeq14T_0$. Gradients are more conspicuous.} 
\end{figure}

Figures \ref{GradX_initial}-\ref{GradX_final} present PDFs for gradients in the 
orthogonal direction. 
\clearpage 
Figures \ref{Eta_surf_initial},\ref{Eta_surf_final} present snapshots of the 
surface in the initial and final moments 
of simulation. Fig.\ref{Eta_surf_max} is a snapshot of the surface in the moment 
of maximal roughness $T=4.94\simeq14T_0$. 
 
\section{Statistical numerical experiment} 
 
\subsection{Numerical model for Hasselmann Equation} 
 
Numerical integration of kinetic equation for gravity waves on deep water 
(Hasselmann equation) was the subject of considerable efforts for last three 
decades. 
The ``ultimate goal'' of the effort -- creation of the operational wave model 
for 
wave forecast based on direct solution of the Hasselmann equation -- happened to 
be an extremely difficult computational problem due to mathematical complexity 
of 
the $S_{nl}$ term, which requires calculation of a three-dimensional integral at 
every advance in time. 
 
Historically, numerical methods of integration of kinetic equation for gravity 
waves exist in two ``flavors''. 
 
The first one is associated with works of \cite{Hasselmann1985}, 
\cite{Dungey1985}, 
\cite{Masuda1981}, \cite{Masuda1986}, \cite{Lavrenov1998} and 
\cite{Polnikov2001}, 
and is based on transformation of 6-fold 
into 3-fold integrals using $\delta$-functions. Such transformation leads to 
appearance of integrable singularities, which creates additional difficulties 
in calculations of the $S_{nl}$ term. 
 
The second type of models has been developed in works of \cite{Webb1978} and 
\cite{Resio1982}, \cite{Resio1991} and is currently known as Resio-Tracy model. 
It uses direct calculation of resonant quadruplet contribution 
into $S_{nl}$ integral, based on the following property: given two fixed vectors 
$\vec{k}, \vec{k_1}$, another two $\vec{k_2}, \vec{k_3}$ are uniquely defined by 
the point ``moving'' along the resonant curve -- locus. 
 
Numerical simulation in the current work was performed with the help of 
modified version of the second type algorithm. Calculations were made on the 
grid $71 \times 36$ points in the frequency-angle domain $[\omega, \theta]$ with 
exponential distribution of points in the frequency domain and uniform 
distribution of points in the angle direction. 
 
To date, Resio-Tracy model suffered rigorous testing and is currently used with 
high degree of trustworthiness: it was tested with respect to  motion integrals 
conservation in the ``clean'' tests, wave action conservation in wave spectrum 
down-shift, realization of self -- similar solution in ``pure swell'' and ``wind 
forced'' 
regimes (see \cite{Pushkarev2000HE}, \cite{Pushkarev2003}, \cite{Badulin2005}). 
 
Description of scaling procedure from dynamical equations to Hasselman kinetic 
equation 
variables is presented in Appedix B. 
 
\subsection{Statistical model setup} 
 
The numerical model used for solution of the Hasselmann equation has been 
supplied with the damping term in three different forms: 
\begin{enumerate} 
\item{} Pseudo-viscous high frequency damping (\ref{Pseudo_Viscous_Damping}) 
used in dynamical equations; 
\item{} $WAM1$ viscous term; 
\item{} $WAM2$ viscous term; 
\end{enumerate} 
Two last viscous terms referred as $WAM1$ and $WAM2$ are the ``white-capping'' 
terms, describing energy 
dissipation by surface waves due to white-capping, as used in $SWAN$ and $WAM$ 
wave forecasting 
models, see \cite{SWAN}: 
\begin{eqnarray} 
\label{WAMdissipation} 
\gamma_{\vec{k}} = C_{ds}  \tilde{\omega} \frac{k}{\tilde{k}} \left((1-\delta)+\
delta\frac{k}{\tilde{k}}\right)\left(\frac{\tilde{S}}{\tilde{S}_{pm}}\right)^p 
\end{eqnarray} 
where $k$ and $\omega$ are wave number and frequency, tilde denotes mean value; 
$C_{ds}$, $\delta$ and $p$ 
are tunable coefficients; $S=\tilde{k}\sqrt{H}$ is the overall steepness; 
$\tilde{S}_{PM}=(3.02\times 10^{-3})^{1/2}$ 
is the value of $\tilde{S}$ for the Pierson-Moscowitz spectrum (note that the 
characteristic steepness 
$\mu = \sqrt{2} S$). 
 
Values of tunable coefficients for $WAM1$ case (corresponding to {\it WAM cycle 
3} dissipation) are: 
\begin{equation} 
C_{ds} = 2.36 \times 10^{-5},\,\,\,\delta=0,\,\,\,p=4 
\end{equation} 
and for $WAM2$ case (corresponding to {\it WAM cycle 4} dissipation) are: 
\begin{equation} 
C_{ds} = 4.10 \times 10^{-5},\,\,\,\delta=0.5,\,\,\,p=4 
\end{equation} 
 
In all three cases we used as initial data smoothed (filtered) spectra 
(Fig.\ref{SmoothedDistribution}) 
obtained in the dynamical run at the time $T_{*} = 3.65 min = 24.3 \simeq 70 
T_0$, which can be considered 
as a moment of the end of the fist "fast" stage of spectral evolution. 
 
The natural question stemming in this point, is why calculation of the dynamical 
and Hasselmann model cannot 
be started from the initial conditions (\ref{Dynamic_initial_conditions}) 
simultaneously? 
 
There are good reasons for that: 
 
As it was mentioned before, the time evolution of the initial conditions 
(\ref{Dynamic_initial_conditions}) in presence of the damping term  can be 
separated in two stages: relatively fast total energy drop in the beginning of 
the evolution and succeeding relatively slow total energy decrease as a function 
of time, see Fig.\ref{H_ArtVisc}. We explain this phenomenon by existence of the 
effective channel of the energy dissipation due to strong nonlinear effects, 
which can be associated with the white-capping. 
 
We have started with relatively steep waves $\mu\simeq 0.167$. As we see, at 
that steepness white-capping is the leading effect. This fact is confirmed by 
numerous field and laboratory experiments. From the mathematical view-point the 
white-capping is formation of coherent structures -- strongly correlated 
multiple harmonics. The spectral peak is posed in our experiments initially at 
$k\simeq 300$, while the edge of the damping area $k_d \simeq 1024$. Hence, only 
the second and the third harmonic can be developed, while hire harmonics are 
suppressed by the strong dissipation. Anyway, even formation of the second and 
the third harmonic is enough to create intensive non-Gaussian tail of the $PDF$ 
for longitudinal gradients. This process is very fast. In the moment of time 
$T=14 T_0$ we see fully developed tails. Relatively sharp gradients mimic 
formation of white caps. Certainly, the ``pure'' Hasselmann equation is not 
applicable on this early stage of spectral evolution, when energy intensively 
dissipates. 
 
As steepness decreases and spectral maximum of the swell down-shifts, an 
efficiency of such mechanism of energy absorption becomes less important when 
the steepness value drops down to $\mu \simeq 0.1$the white-capping becomes a 
negligibly small effect. It happens at $T\simeq 280 T_0$. We decided to start 
comparison between deterministic and statistical modeling in some intermediate 
moment of time $T\simeq 70 T_0$. 
 
\section{Comparison of deterministic and statistical experiments.} 
 
\subsection{Statistical experiment with pseudo-viscous damping term.} 
 
First simulation has been performed with pseudo-viscous damping term, equivalent 
to (\ref{Pseudo_Viscous_Damping}). 
 
Fig.\ref{N_ArtVisc} -- \ref{MeanFreq_ArtVisc} show total wave action, total 
energy, mean wave slope and mean wave frequency as the functions of time. 
 
\begin{figure}[ht]
\centering 
\includegraphics[width=5.0cm,angle=90]{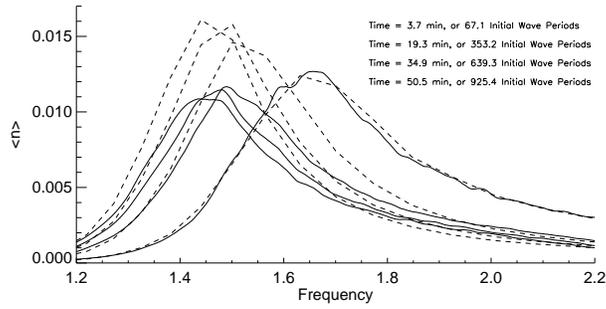} 
\caption{Angle-averaged spectrum as a function of time for dynamical and Hasselmann equations for artificial viscosity case.}\label{AngleAver_ArtVisc} 
\end{figure}

Fig.\ref{AngleAver_ArtVisc} shows the time evolution of angle-averaged wave 
action spectra as the functions of frequency for dynamical and Hasselmann 
equations. 
 
Temporal behavior of angle-averaged spectrum is presented on 
Fig.\ref{AngleAver_ArtVisc}. We see the down-shift of the spectral maximum both 
in dynamic and Hasselmann equations. The correspondence of the spectral maxima 
is not good at all. 
 
It is obvious that the influence of the artificial viscosity is not strong 
enough to reach  the correspondence of two models. 
 
\subsection{Statistical experiments with $WAM1$ damping term} 
 
\begin{figure}[ht]
\centering 
\includegraphics[scale=0.4]{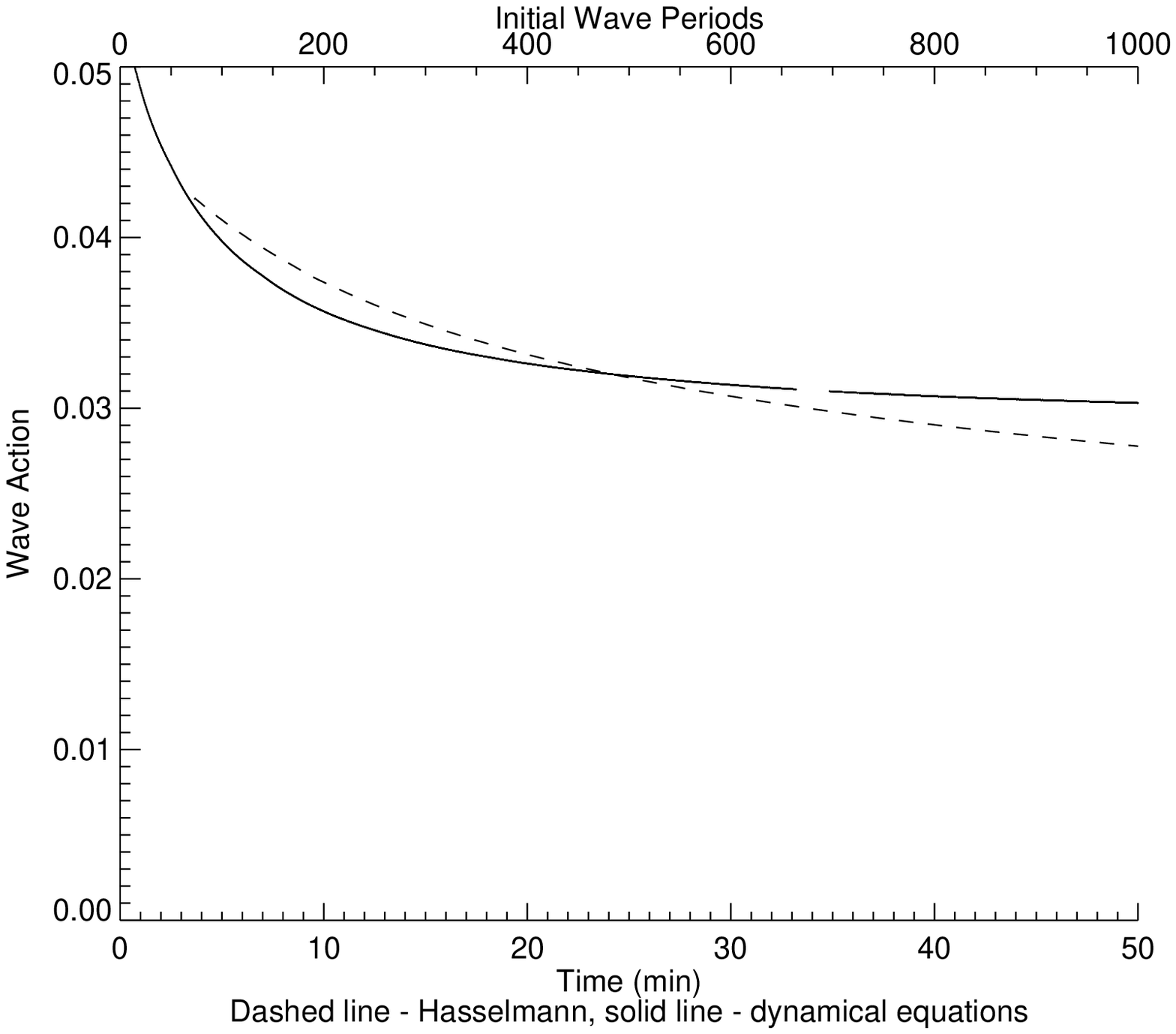} 
\caption{Total wave action as a function of time for $WAM1$ case.}\label{N_WAM2} 
\end{figure}

\begin{figure}[ht]
\centering 
\includegraphics[scale=0.4]{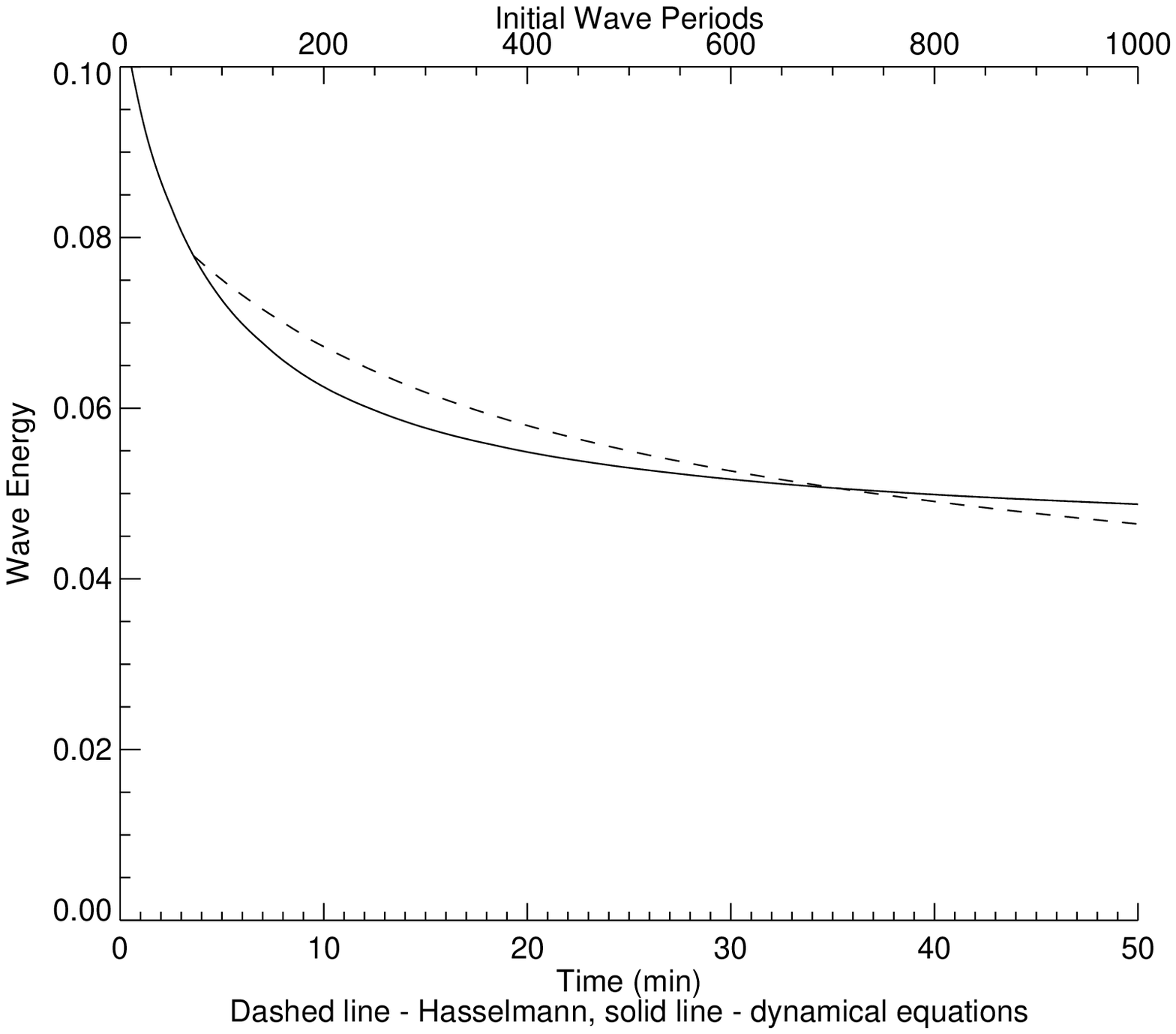} 
\caption{Total wave energy as a function of time for $WAM1$ case}\label{H_WAM2} 
\end{figure}

\begin{figure}[ht]
\centering 
\includegraphics[scale=0.4]{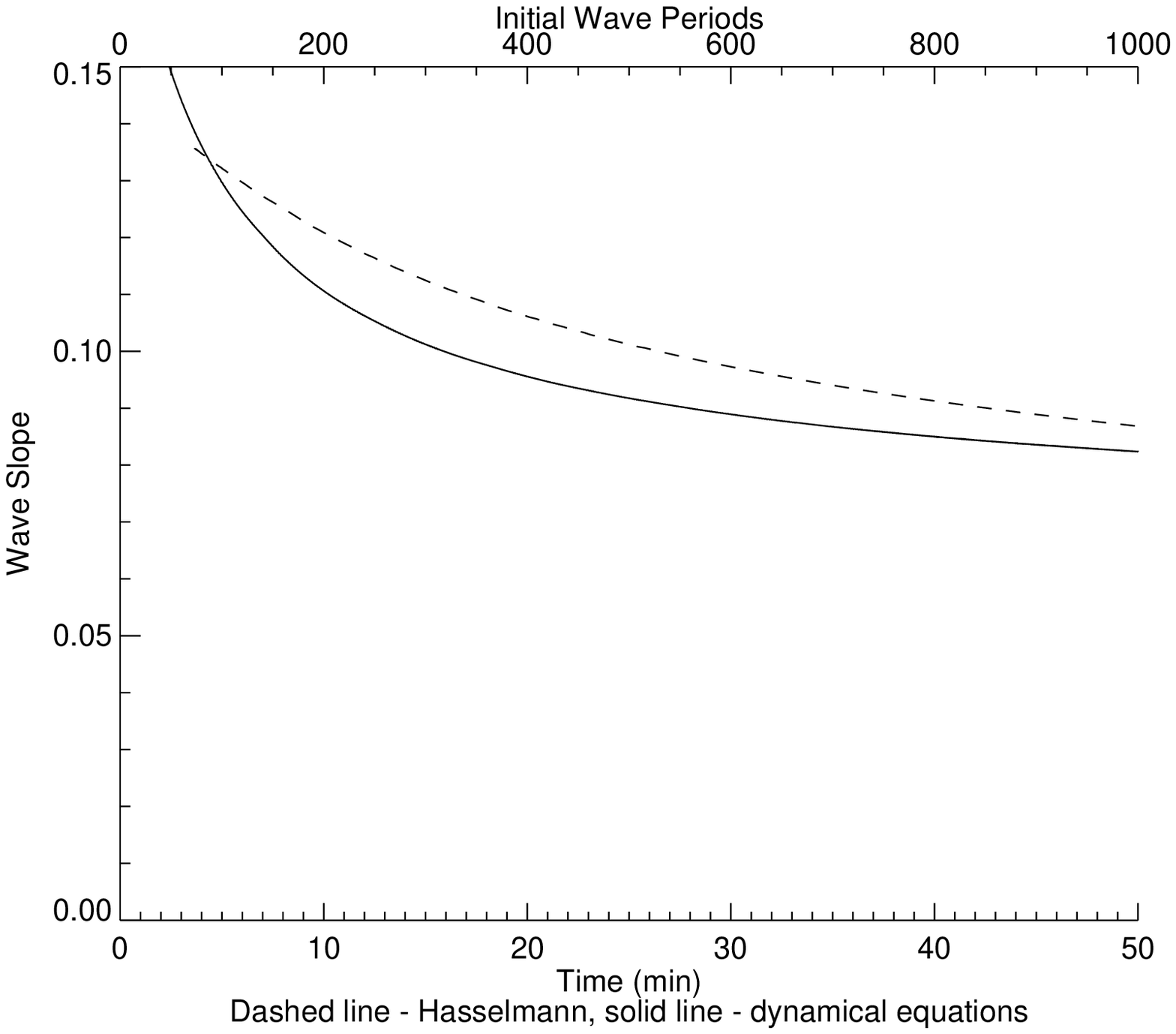} 
\caption{Average wave slope as a function of time for $WAM1$ case.}\label{Slope_WAM2} 
\end{figure}

\begin{figure}[ht]
\centering 
\includegraphics[scale=0.4]{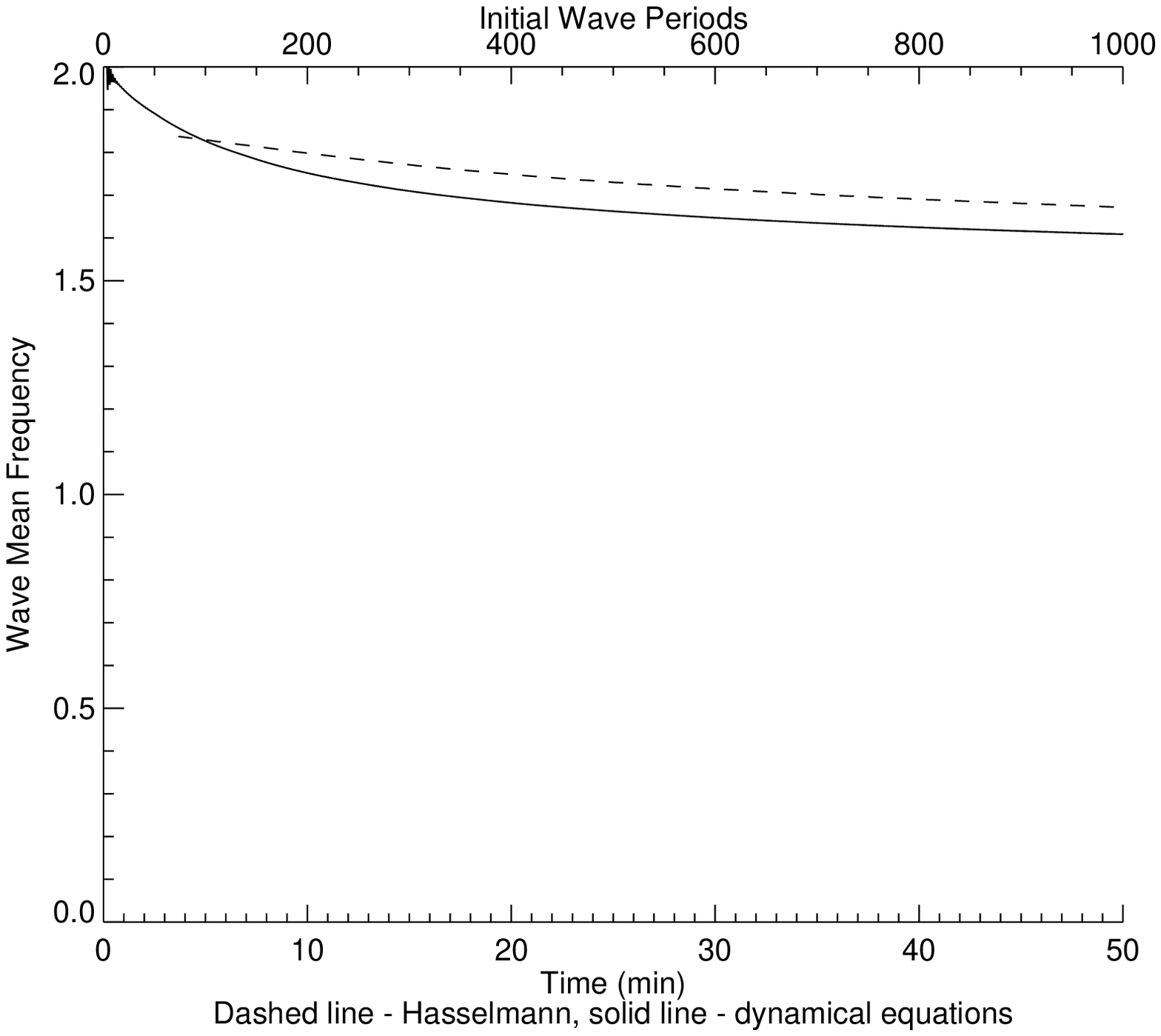} 
\caption{Mean wave frequency as a function of time for $WAM1$ case.}\label{MeanFreq_WAM2} 
\end{figure}

Fig.\ref{N_WAM2} -- \ref{MeanFreq_WAM2} show total wave action, total energy, 
mean wave slope and mean wave frequency as the functions of time. 
 
The temporal behavior of total wave action, energy and average wave slope is 
much better than in the artificial viscosity term, and for $50 \, min$ duration 
of the experiment  we observe decent correspondence between dynamical and 
Hasselmann equations. However for longer time the $WAM1$ model deviates from the
exact calculations significantly.
 
It is important to note that the curves of temporal behavior of the total wave 
action, energy and average wave slope diverge at the end of simulation time with 
different derivatives, and the correspondence cannot be expected to be that good 
outside of the simulation time interval. 
 
\begin{figure}[ht]
\centering 
\includegraphics[width=4.5cm,angle=90]{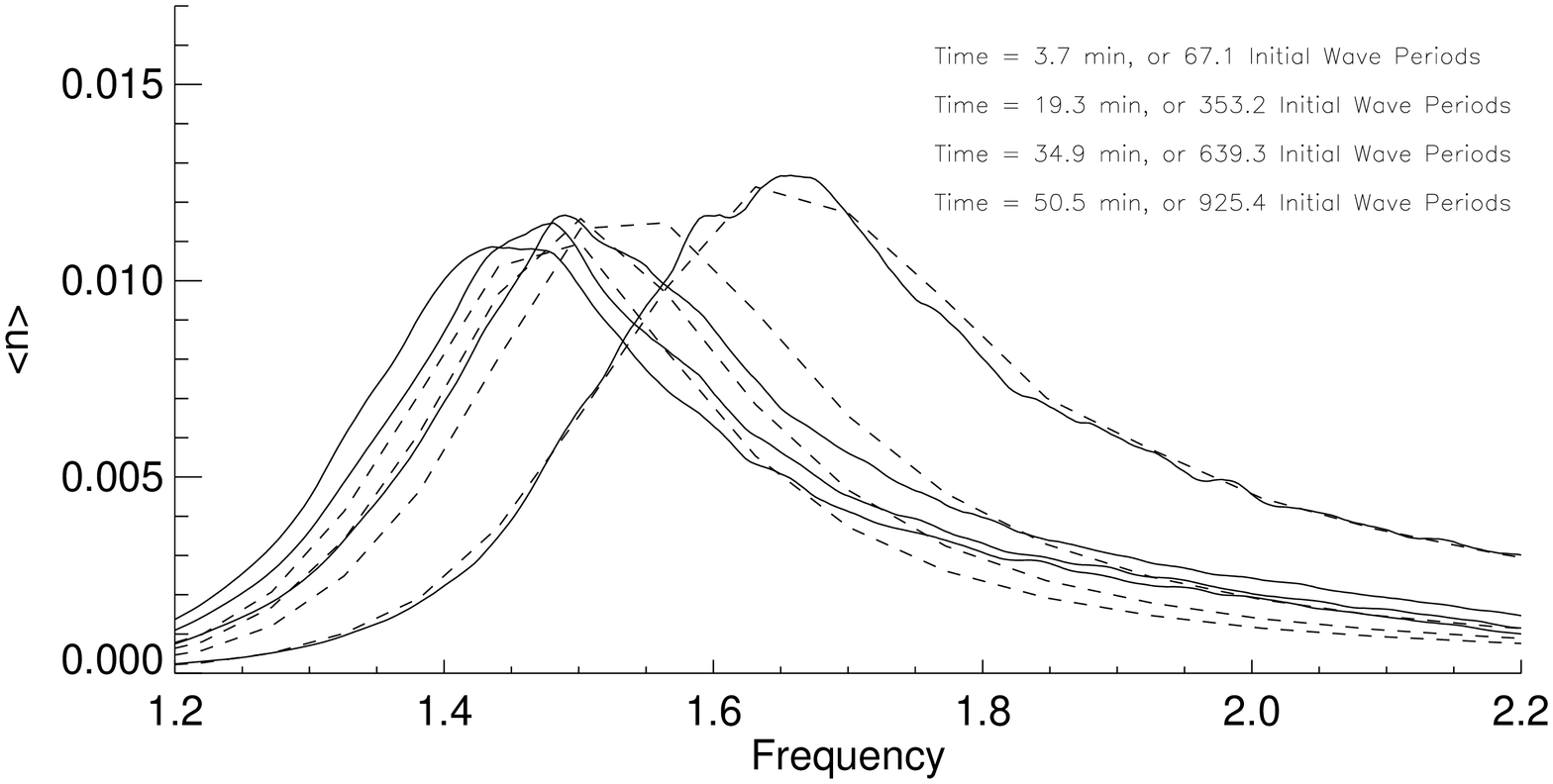} 
\caption{Angle-averaged spectrum as a function of time for dynamical and Hasselmann equations a function of time for $WAM1$ case.}\label{AngleAver_WAM2} 
\end{figure}

Fig.\ref{AngleAver_WAM2} shows the time evolution of the angle-averaged wave 
action spectra as the functions of frequency for dynamical and Hasselmann 
equations. As in the artificial viscosity case, we observe distinct down-shift 
of the spectral maxima. Correspondence of the time evolution of the amplitudes 
of the spectral maxima is much better then in artificial viscosity case.

\subsection{Statistical experiments with $WAM2$ damping term} 
 
\begin{figure}[ht]
\centering 
\includegraphics[scale=0.4]{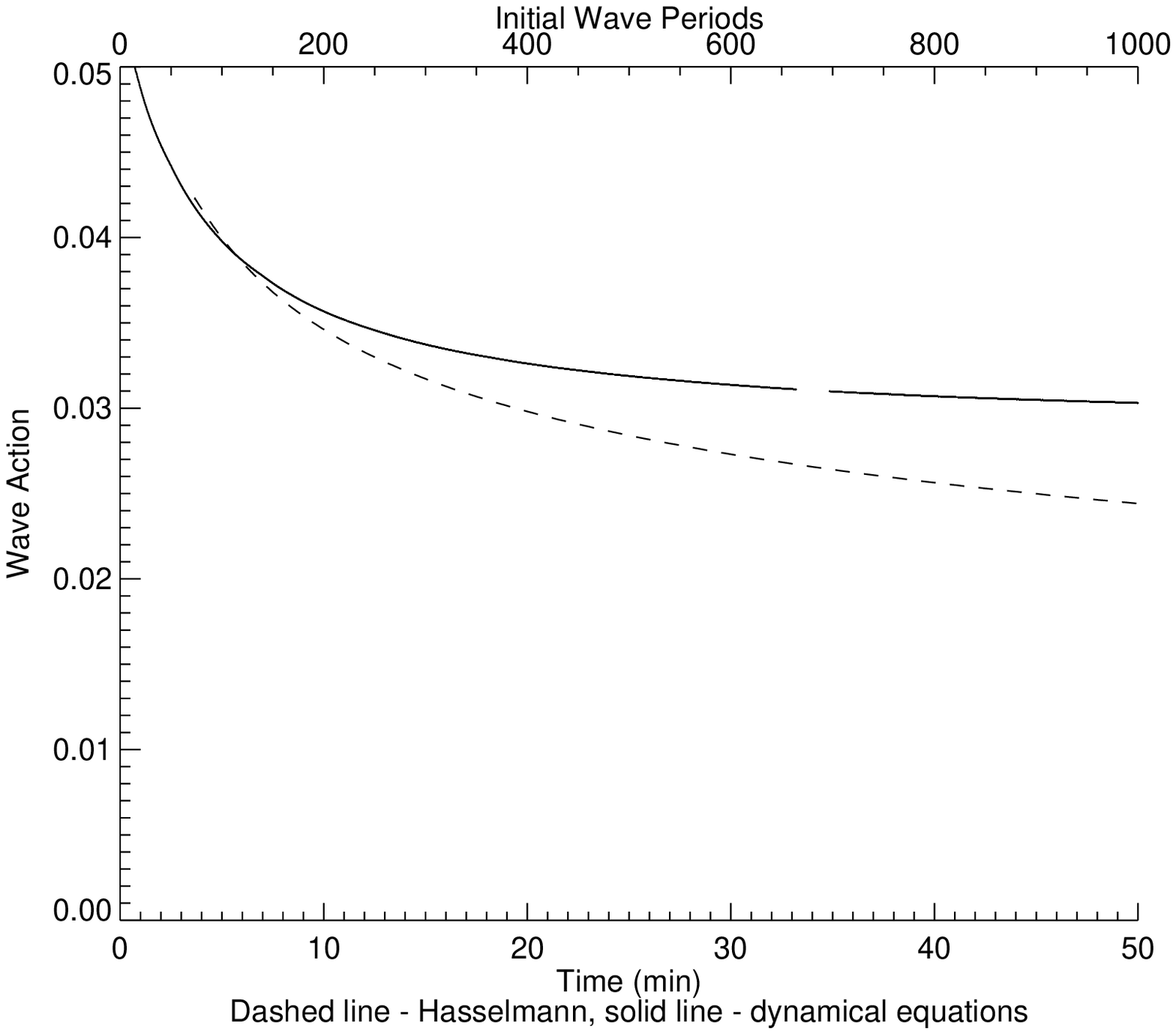} 
\caption{Total wave action as a function of time for $WAM2$ case.}\label{N_WAM1} 
\end{figure}

\begin{figure}[ht]
\centering 
\includegraphics[scale=0.4]{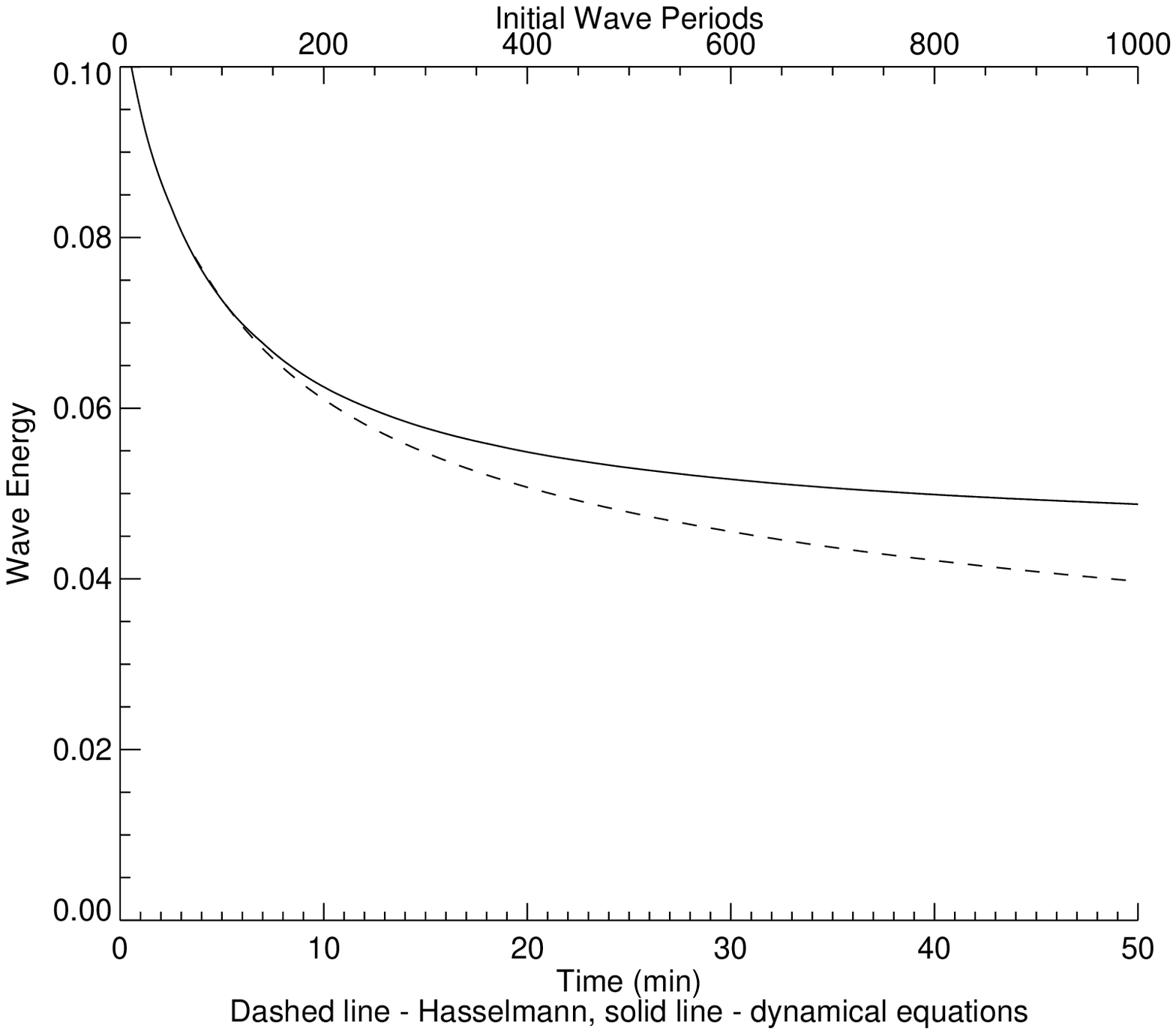} 
\caption{Total wave energy as a function of time for $WAM2$ case}\label{H_WAM1} 
\end{figure}

\begin{figure}[ht]
\centering 
\includegraphics[scale=0.4]{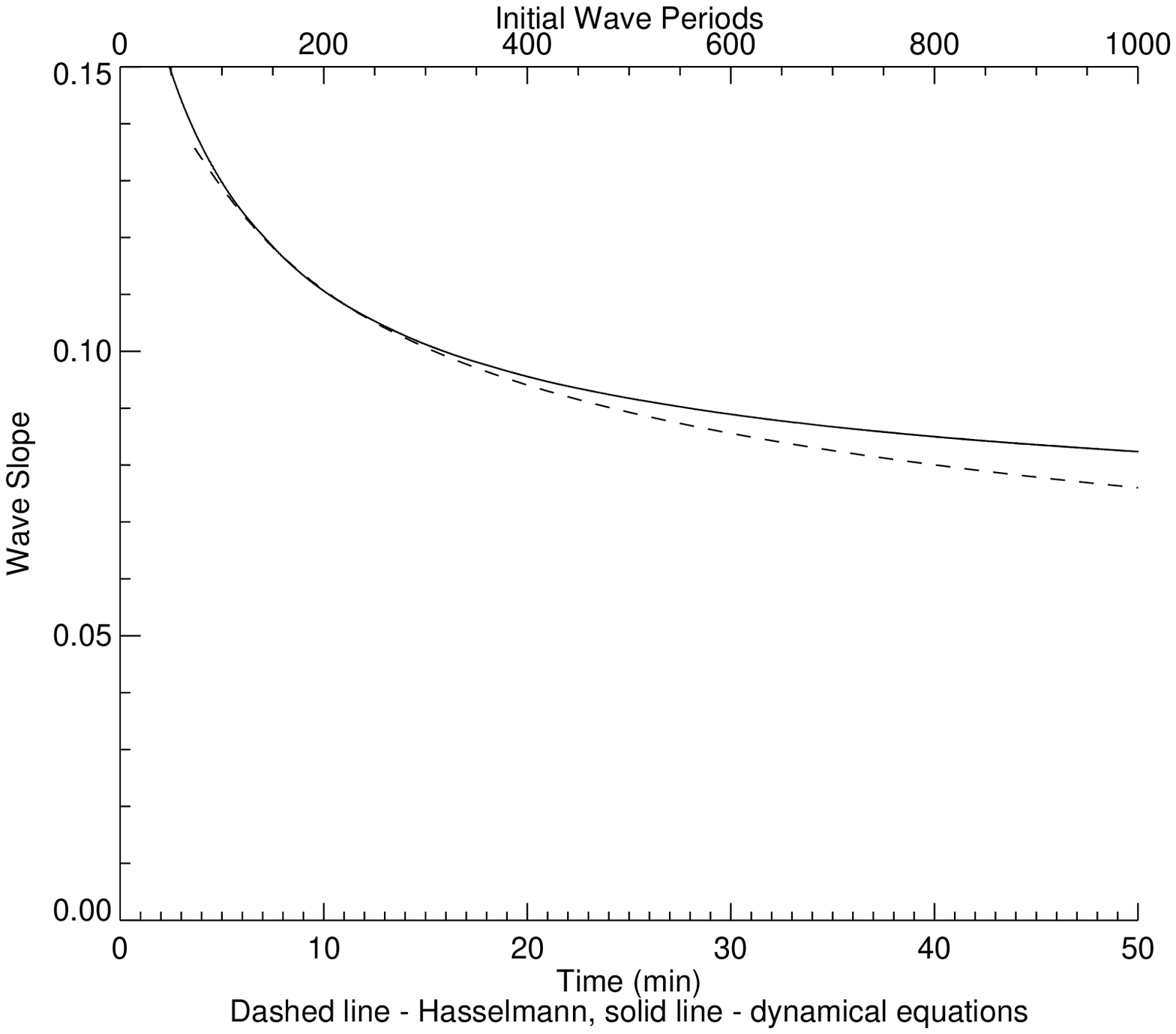} 
\caption{Average wave slope as a function of time for $WAM2$ 
case.}\label{Slope_WAM1} 
\end{figure}

\begin{figure}[ht]
\centering 
\includegraphics[scale=0.4]{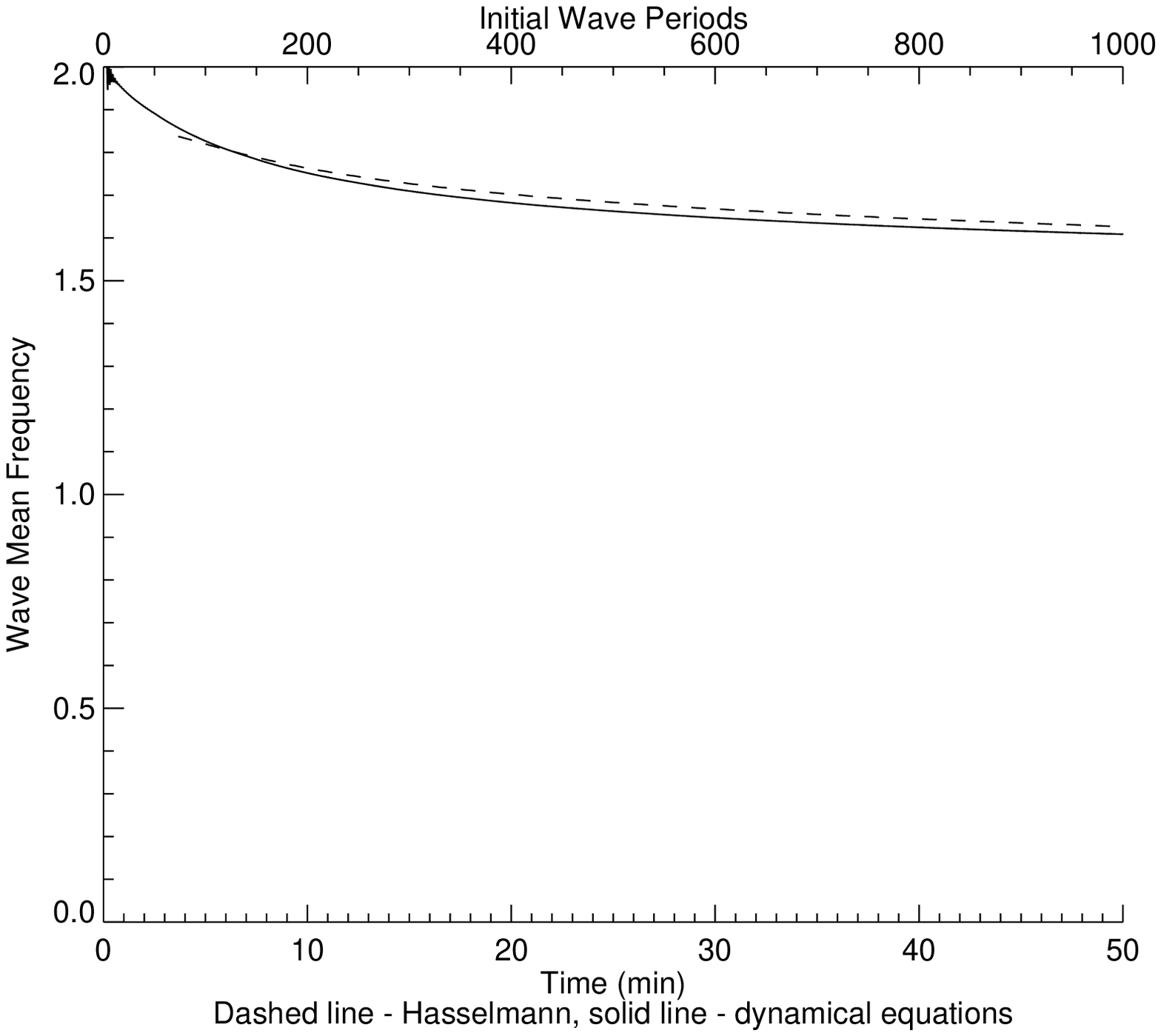} 
\caption{Mean wave frequency as a function of time for $WAM2$ 
case.}\label{MeanFreq_WAM1} 
\end{figure}

Fig.\ref{N_WAM1} -- \ref{MeanFreq_WAM1} shows the temporal evolution of the 
total wave action, total energy, mean wave slope and mean wave frequency, which 
are divergent in time. 
 
\begin{figure}[ht]
\centering 
\includegraphics[width=5.0cm,angle=90]{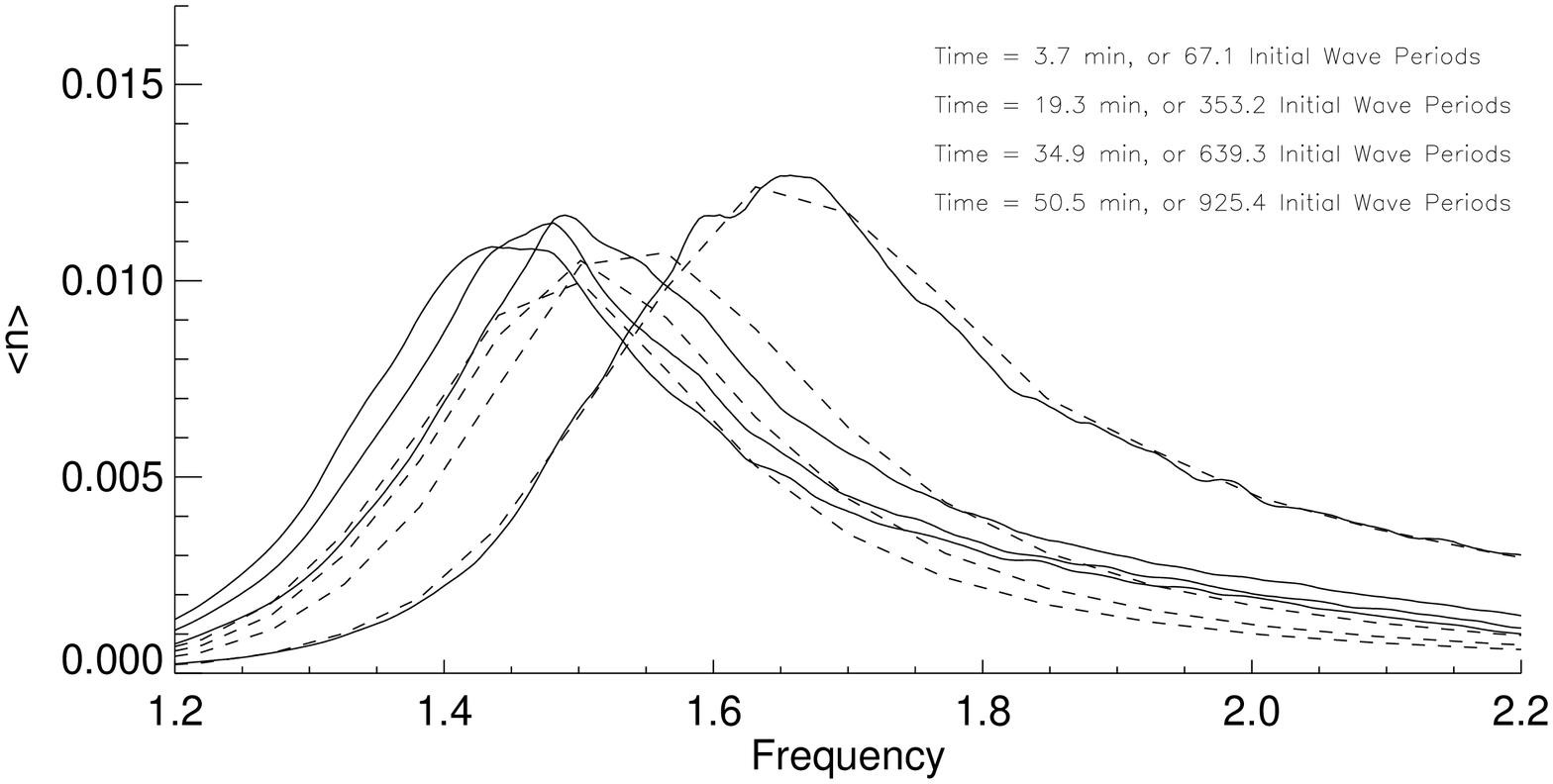} 
\caption{Angle-averaged spectrum as a function of time for dynamical and Hasselmann equations a function of time for $WAM2$ case.}\label{AngleAver_WAM1} 
\end{figure}

Fig.\ref{AngleAver_WAM1} show time evolution of angle-averaged wave action 
spectra as the functions of frequency for dynamical and Hasselmann equations. 
While as in the artificial viscosity and $WAM1$ cases we also observe distinct 
down-shift of the spectral maxima, the correspondence of the time evolution of 
the amplitudes of the spectral maxima is worse than in $WAM1$ case. 
 
Despite the fact that historically $WAM2$ appeared as an improvement of $WAM1$ 
damping term, it does not improve the correspondence of two models, observed in 
$WAM1$ case, and is presumably too strong for description of the reality.

\section{Down-shift and angular spreading} 
 
The major process of time-evolution of the swell is frequency down-shift. During 
$T=933 T_0$ the mean frequency has been decreased from $\omega_0 = 2$ to 
$\omega_1 = .6$. On the last stage of the process, the mean frequency slowly 
decays as 
 
\begin{equation} 
\label{Law1} 
<\omega> \simeq t^{-0.067} \simeq t^{-1/15} 
\end{equation} 
 
The Hasselmann equation has self-similar solution, describing the evolution of 
the swell $n(\vec{k},t) \simeq t^{4/11} F\left(\frac{\vec{k}}{t^{2/11}}\right)$ 
(see \cite{Pushkarev2003}, \cite{Badulin2005}). For this solution 
\begin{equation} 
\label{Law2} 
<\omega> \simeq t^{-1/11} 
\end{equation} 
 
\begin{figure}[htb!]
\centering 
\includegraphics[width=8.5cm,angle=0]{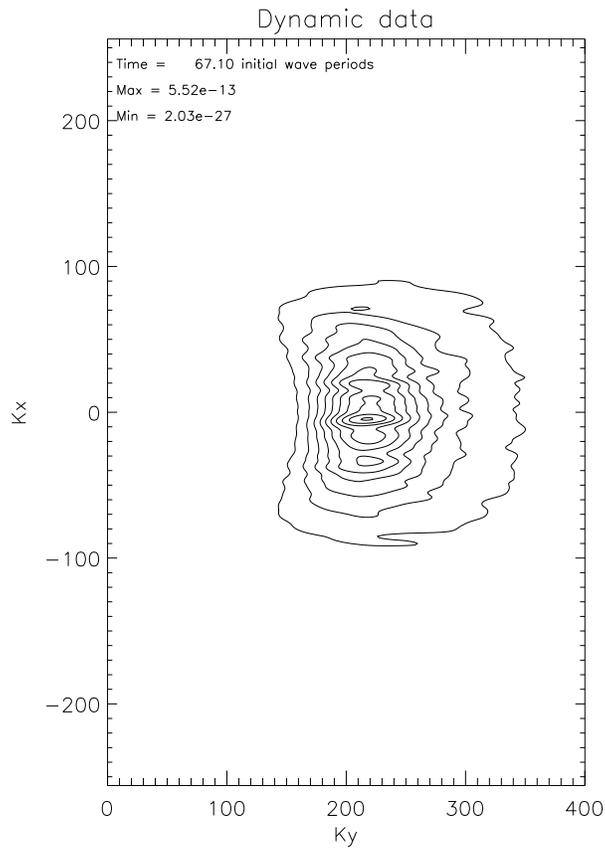} 
\caption{\label{SpreadDyn1}Level lines of the spectra at $t = 67T_0$. Dynamical equations.}
\end{figure}

\begin{figure}[htb!]
\centering 
\includegraphics[width=8.5cm,angle=0]{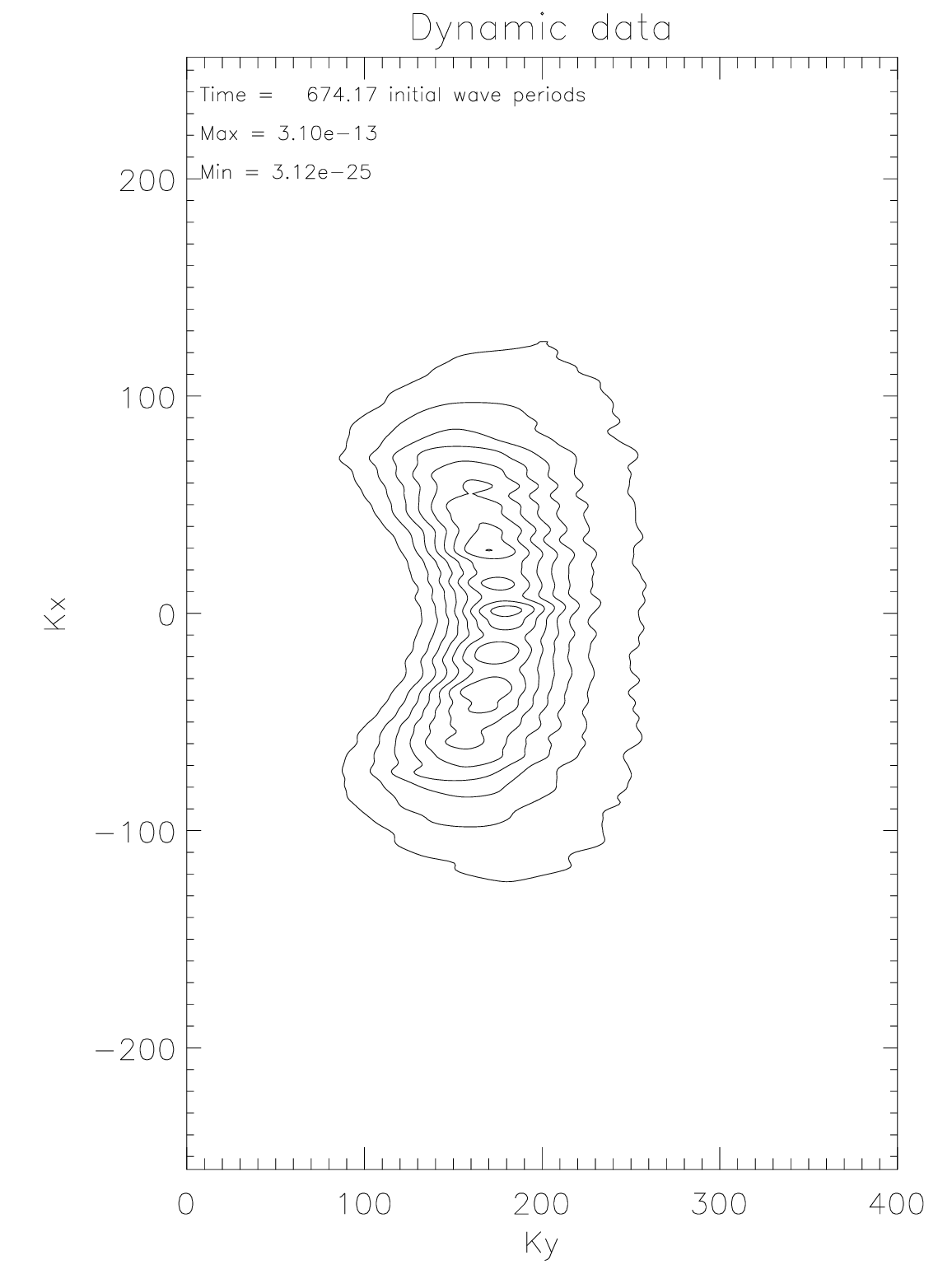} 
\caption{\label{SpreadDyn2}Level lines of the spectra at $t = 674T_0$. Dynamical equations.}
\end{figure}

\begin{figure}[htb!]
\centering 
\includegraphics[width=8.5cm,angle=0]{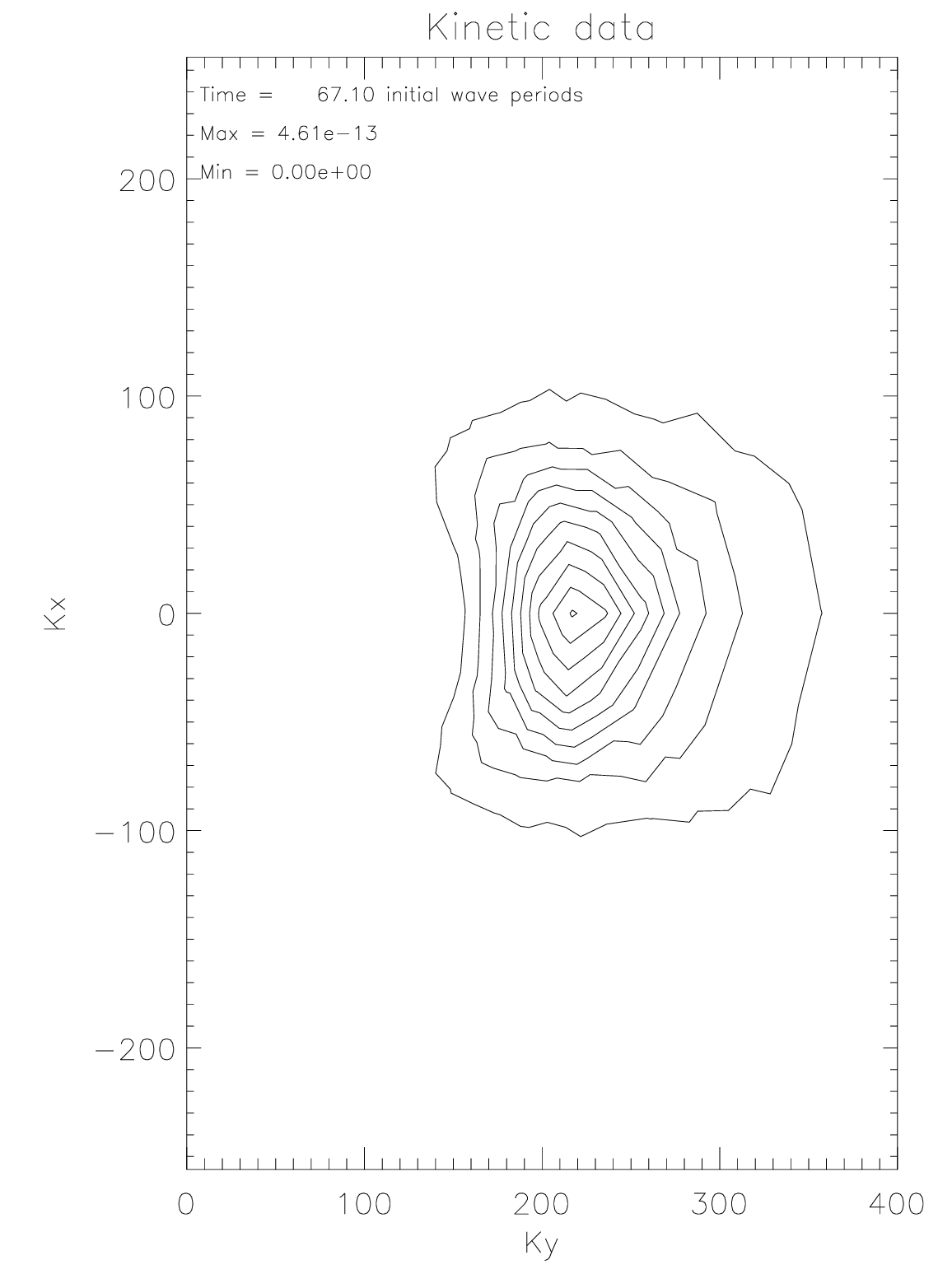} 
\caption{\label{SpreadHass1}Level lines of the spectra at $t = 67T_0$. Hasselmann equation.}
\end{figure}

\begin{figure}[htb!]
\centering 
\includegraphics[width=8.5cm,angle=0]{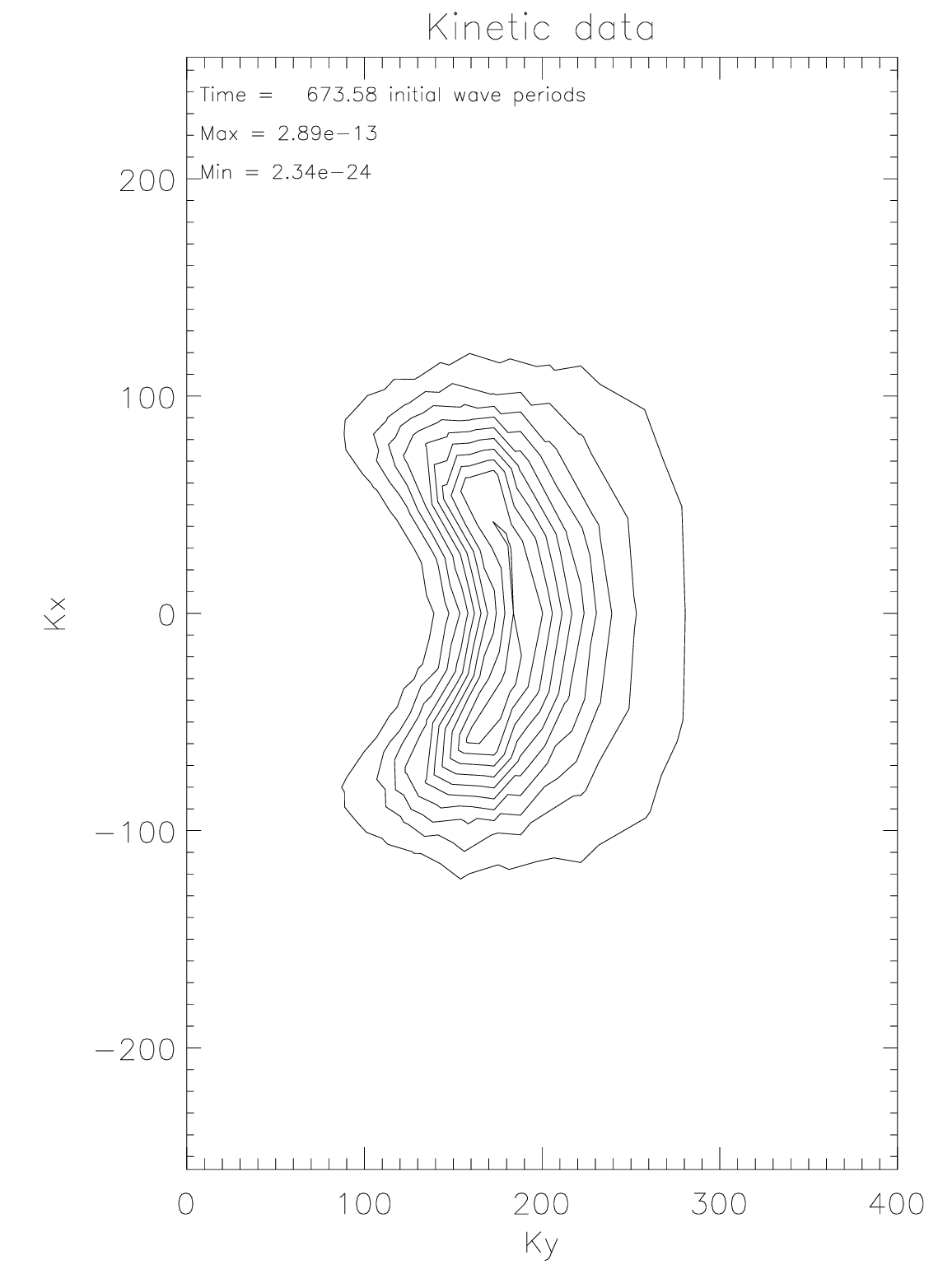} 
\caption{\label{SpreadHass2}Level lines of the spectra at $t = 674T_0$. Hasselmann equation.}
\end{figure}

\begin{figure}[htb!]
\centering 
\includegraphics[width=8.5cm,angle=0]{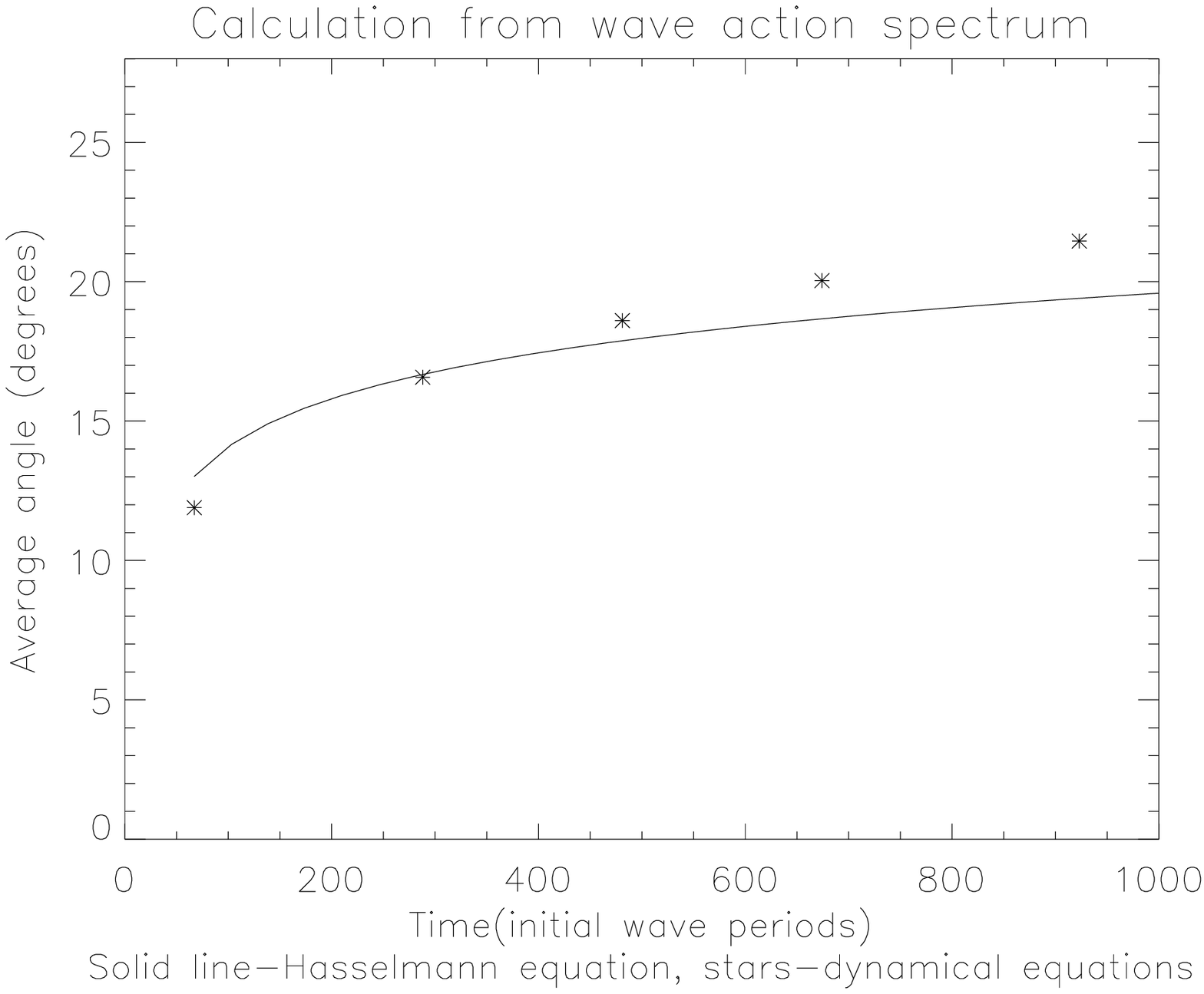} 
\caption{\label{SpreadCompareAction}Comparison of time-evolution of the mean angular spreading
$\left(\int |\theta| n(\vec k) \D\vec{k}\right)/\left(\int n(\vec k) \D\vec{k}\right)$ calculated through wave action spectra.}
\end{figure}

\begin{figure}[htb!]
\centering 
\includegraphics[width=8.5cm,angle=0]{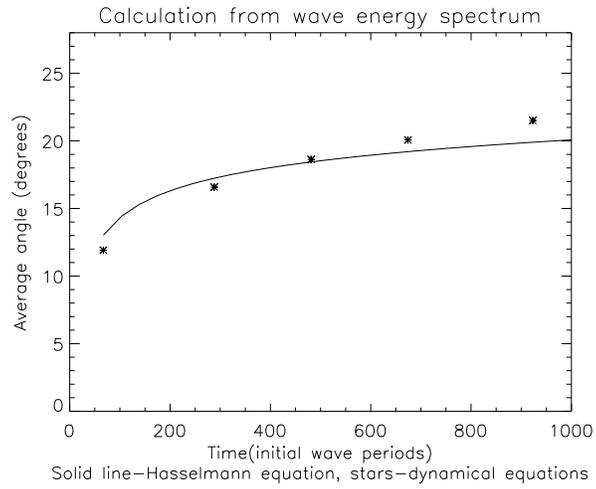} 
\caption{\label{SpreadCompareEnergy}Comparison of time-evolution of the mean angular spreading
$\left(\int |\theta| \omega n(\vec k) \D\vec{k}\right)/\left(\int \omega n(\vec k) \D\vec{k}\right)$ calculated through wave energy spectra.}
\end{figure}

The difference between (\ref{Law1}) and (\ref{Law2}) can be explained as 
follows. What we observed, is not a self-similar behavior. Indeed, a 
self-silmilarity presumes that the angular structure of the solution is constant 
in time. Meanwhile, we observed intensive angular spreading of the initially 
narrow in angle, almost one-dimensional wave spectrum.
Level lines of the spectra after low-pass filtering, obtained in dynamical
equations simulation,
for two moments of time are presented on Fig. \ref{SpreadDyn1}-\ref{SpreadDyn2}.
Level lines of the spectra in the same moments of time, obtained by solution
of the Hasselmann equation are presented on Fig. \ref{SpreadHass1}-\ref{SpreadHass2}.
One can see good correspondance between results of both experiments. Comparison
of time-evolution of the mean angular spreading calculated from action and energy
spectra are presented on Fig. \ref{SpreadCompareAction}-\ref{SpreadCompareEnergy}.
\clearpage
One has
to expect that the angular spreading will be arrested at later times, and the spectra
will take a universal self-similar shape.

\section{Conclusion} 
 
1. We started our experiment with characteristic steepness $\mu \simeq 0.167$. 
This is three times less than steepness of the Stokes 
wave of limiting amplitude, but still it is a large steepness typical for young 
waves. For waves of such steepness white-capping 
effect could be essential. However, in our experiments we cannot observe such 
effects due to the strong pseudo-viscosity. 
Indeed, third harmonics of our initial leading wave is situated near the edge of 
damping area, while fourth and higher harmonics 
are far in the damping area. This circumstanse provides an intensive energy 
dissipation, which is not described by the Hasselmann equation. 
 
Anyway, on the first stage of the process we observe intensive generation of 
coherent higher harmonics which reveal itself in 
tails of PDF for longitudinal gradients. If our damping region would be shifted 
further to higher wave numbers, we could 
observe sharp crests formation. 
 
2. We ended up with steepness $\mu \simeq 0.09$. This is close to mature waves, 
typically observed in the ocean and described 
by Hasselmann equation pretty well. We observed characteristic effects predicted 
by the weak-turbulent theory --- down-shift 
of mean frequency formation, Zakharov-Filonenko weak turbulent spectrum 
$\omega^{-4}$ and strong angular spreading. Comparison 
of time-derivatives of the average quantities shows that for this steepness 
wave-breaking (white-capping) become not essential at $\mu \simeq 0.09$. 
 
In general, our experiments validate Hasselmann equation. However, it has to be 
accomplished by a proper dissipation term. 
 
3. The dissipative term used in the $WAM1$ model fairly describe damping due to 
white capping on the initial stage of evolution. 
It overestimate damping, however, for moderate steepness $\mu \simeq 0.09$ 
 
The dissipative term, used in the $WAM2$ model is not good. It overestimates 
damping essentially. 
 
4. Presence of abnormally intensive harmonics, so called "oligarchs" show that, 
in spite of using a very fine grid, we did not 
eliminated effects of discreteness completely. More accurate verification of the 
Hasselmann equation should be made on the grid 
containing more than $10^7$ modes. This is quite realistic task for modern 
supercomputers, and we hope to realize 
such an experiment. 
 
Another conclusion is more pessimistic. Our results show that it is very 
difficult to reproduce real ocean conditions in 
any laboratory wave tank. Even a tank of size $200\times200$ meters is not large 
enough to model ocean due to the presence 
of wave numbers grid discreteness. 
 
\section{Acknowledgments} 
This work was 
supported by ONR grant N00014-03-1-0648, RFBR grant 06-01-00665-a, INTAS grant 
00-292, the Programme 
``Nonlinear dynamics and solitons'' from the RAS Presidium and ``Leading 
Scientific 
Schools of Russia" grant, also by US Army Corps of Engineers 
Grant DACW 42-03-C-0019 and by NSF Grant NDMS0072803.

A.O. Korotkevich was supported by Russian President grant for young scientist MK-1055.2005.2.
 
Also authors want to thank the creators of the open-source fast Fourier 
transform library 
FFTW~\cite{FFTW} for this fast, portable and completely free piece of software. 

\section{Appendix A. "Forbes list of 15 largest harmonics.} 
Here one can find 15 largest harmonics at the end of calculations in the 
framework of dynamical equations. 
Average square of amplitudes in dissipation-less region was taken from smoothed 
spectrum, while 
all these harmonics exceed level $|a_{\vec k}|^2 = 1.4\times10^{-12}$. 

\begin{tabular}{|c|c|c|c|c|}
\hline
$K_x$ & $K_y$ & $|a_{\vec k}|^2$ & $<|a_{\vec k}|^2>_{filter}$ & $|a_{\vec k}|^2/<|a_{\vec k}|^2>$\\
\hline
-59 & 155 & 1.563e-12 & 0.746e-13 & 2.095e+1 \\
\hline
-37 & 166 & 1.903e-12 & 1.201e-13 & 1.585e+1 \\
\hline
-37 & 185 & 1.569e-12 & 2.288e-13 & 0.686e+1 \\
\hline
-36 & 162 & 1.477e-12 & 0.992e-13 & 1.489e+1 \\
\hline
-33 & 157 & 1.442e-12 & 0.713e-13 & 2.022e+1 \\
\hline
-26 & 164 & 3.351e-12 & 0.847e-13 & 3.956e+1 \\
\hline
-17 & 189 & 1.463e-12 & 2.789e-13 & 0.525e+1 \\
\hline
-14 & 173 & 1.408e-12 & 1.459e-13 & 0.965e+1 \\
\hline
-2 & 176 & 1.533e-12 & 1.697e-13 & 0.903e+1 \\
\hline
0 & 177 & 2.066e-12 & 1.741e-13 & 1.187e+1 \\
\hline
10 & 179 & 1.427e-12 & 1.893e-13 & 0.754e+1 \\
\hline
27 & 163 & 1.483e-12 & 0.832e-13 & 1.782e+1 \\
\hline
31 & 174 & 1.431e-12 & 1.342e-13 & 1.066e+1 \\
\hline
37 & 173 & 1.578e-12 & 1.581e-13 & 0.998e+1 \\
\hline
60 & 133 & 1.565e-12 & 0.345e-13 & 4.536e+1 \\
\hline
\end{tabular}

\subsection{Appendix B. From Dynamical Equations to \\ Hasselmann Equation.} 
Standard setup for numerical simulation of the dynamical equations 
(\ref{eta_psi_equations}), implies $2 \pi \times 2 \pi$ 
domain in real space and gravity acceleration $g=1$. Usage of the domain size 
equal $2 \pi$ is convenient because in this case 
wave numbers are integers. 
 
In the contrary to dynamical equations, the kinetic equation 
(\ref{Hasselmann_equation}) is formulated in terms of real 
physical variables and it is necessary to describe the transformation from the 
``dynamical'' variables into to 
the ``physical'' ones. 
 
Eq.\ref{eta_psi_equations} are invariant with respect to ``stretching'' 
transformation from ``dynamical'' to ``real'' variables: 
\begin{eqnarray} 
\eta_{\vec{r}} &=& \alpha \eta_{\vec{r}^\prime}^\prime,\,\,\,\,\vec{k} = 
\frac{1}{\alpha} \vec{k}^\prime,\,\,\,\,\vec{r} = \alpha 
\vec{r}^\prime,\,\,\,\,g = \nu g^\prime, \\ 
t &=& \sqrt{\frac{\alpha}{\nu}}t^\prime,\,\,\, L_x = \alpha L_x^\prime,\,\,\, 
L_y = \alpha L_y^\prime 
\end{eqnarray} 
where prime denotes variables corresponding to dynamical equations. 
 
In current simulation we used the stretching coefficient $\alpha=800.00$, which 
allows to reformulate the statement of the 
problem in terms of real physics: we considered $5026\,m \times 5026\,m$ 
periodic boundary conditions domain of 
statistically uniform ocean with the same resolution in both directions and 
characteristic wave length of the initial 
condition around $22\, m$. In oceanographic terms, this statement corresponds to 
the ``duration-limited 
experiment''.

\printindex
\end{document}